\documentclass[a4paper,12pt]{article}
\usepackage[english]{babel}
\usepackage{amsmath,amsfonts,amssymb}
\usepackage{graphicx}

\usepackage{algorithm,algpseudocode}

\usepackage{pgfplots}
\usepackage{amsthm}
\usepackage{hyperref}
\newcommand{\C}{\mathbb C}

\newcommand{\rmd}{\mathrm d}
\newcommand{\rmi}{\mathrm i}
\DeclareMathOperator{\ad}{ad}
\newcommand{\Om}{\Omega}
\DeclareMathOperator{\dexp}{dexp}

\theoremstyle{definition}

\newtheorem{bsp}{Example}
\newtheorem{met}{Method}
\theoremstyle{remark}

\pgfplotsset{height=0.4\linewidth,width=0.96\linewidth,compat=1.10,
    every axis/.append style={legend style={
        /tikz/every even column/.append style={column sep=9pt}}}}

\definecolor{mblue}{RGB}{0,74,255}
\definecolor{mgreen}{RGB}{0,200,0}
\definecolor{morange}{RGB}{255,147,49}
\definecolor{mred}{RGB}{255,0,0}

\pgfplotscreateplotcyclelist{short_list}{%
mblue,line width=1.3pt, mark=*, solid\\
mgreen,line width=1.3pt, mark=square*, solid\\
morange,line width=1.3pt, mark=triangle*, solid\\
brown,line width=1.3pt, mark=diamond*, solid\\
mred,line width=1.3pt, mark=star, solid\\
gray, line width=1pt, mark=none, dashed\\%
gray, line width=1pt, mark=none, dotted\\%
gray, line width=1pt, mark=none, dashdotted\\%
} 

\pgfplotscreateplotcyclelist{long_list}{%
mblue,line width=1.3pt, mark=*, solid\\
mgreen,line width=1.3pt, mark=square*, solid\\
morange,line width=1.3pt, mark=triangle*, solid\\
mblue,line width=1.3pt, mark=*, dashed\\
mgreen,line width=1.3pt, mark=square*, dashed\\
morange,line width=1.3pt, mark=triangle*, dashed\\
} 

\pgfplotscreateplotcyclelist{long_list2}{%
mblue,line width=1.3pt, mark=*, solid\\
mgreen,line width=1.3pt, mark=square*, solid\\
morange,line width=1.3pt, mark=triangle*, solid\\
brown,line width=1.3pt, mark=diamond*, solid\\
mred,line width=1.3pt, mark=star, solid\\
mblue,line width=1.3pt, mark=*, dashed\\
mgreen,line width=1.3pt, mark=square*, dashed\\
morange,line width=1.3pt, mark=triangle*, dashed\\
brown,line width=1.3pt, mark=diamond*, dashed\\
mred,line width=1.3pt, mark=star, dashed\\
mblue,line width=1.3pt, mark=*, dotted\\
mgreen,line width=1.3pt, mark=square*, dotted\\
morange,line width=1.3pt, mark=triangle*, dotted\\
brown,line width=1.3pt, mark=diamond*, dotted\\
mred,line width=1.3pt, mark=star, dotted\\
} 

\usepgfplotslibrary{groupplots}
\usetikzlibrary{external}

\pgfkeys{/pgf/images/include external/.code={\includegraphics{#1}}}

\newenvironment{customlegend}[1][]{%
    \begingroup
    \csname pgfplots@init@cleared@structures\endcsname
    \pgfplotsset{#1}%
}{%
    \csname pgfplots@createlegend\endcsname
    \endgroup
}%
\def\addlegendimage{\csname pgfplots@addlegendimage\endcsname}

\newcounter{mylineno}
\makeatletter
\let\oldtabcr\@tabcr

\def\mynewline{\refstepcounter{mylineno}%
               \llap{\footnotesize\arabic{mylineno}\hspace{5pt}}%
              }

\gdef\@tabcr{\@stopline \@ifstar{\penalty%
           \@M \@xtabcr}\@xtabcr\mynewline}

\makeatother

\usepackage[numbers, sort]{natbib}
\bibpunct{[}{]}{,}{n}{}{;} 

\usepackage{hyperref}
\hypersetup{
    pdfsubject={Master's thesis},
    pdfcreator={LaTeX2e},
    colorlinks=true,
    linkcolor=blue,
    citecolor=blue,
    filecolor=magenta,
    urlcolor=blue!50!black
}

\newcommand{\includegraphicstikz}[1]{
\IfFileExists{./#1.pdf}{}{
	\immediate\write18{pdflatex -halt-on-error -interaction=batchmode -jobname "#1" "\string\def\string\tikzexternalrealjob\string{\jobname-images\string}\string\input\string{\jobname-images\string}"}
}
\includegraphics{#1}
}

\title{Magnus integrators on multicore CPUs and GPUs}
\author{N.\ Auer, L.\ Einkemmer, P.\ Kandolf, and A. Ostermann \\
                            Department of Mathematics, University of Innsbruck, Austria}
\begin{document}
\maketitle
\begin{abstract}

    In the present paper we consider numerical methods to solve the \textcolor{black}{discrete} Schr\"odinger equation with a time dependent Hamiltonian \textcolor{black}{(motivated by problems encountered in the study of spin systems).} We will consider both short-range interactions, which lead to evolution equations involving sparse matrices, and long-range interactions, which lead to dense matrices. Both of these settings show very different computational characteristics. We use Magnus integrators for time integration and employ a framework based on Leja interpolation to compute the resulting action of the matrix exponential. We consider both traditional Magnus integrators (which are extensively used for these types of problems in the literature) as well as the recently developed commutator-free Magnus integrators and implement them on modern CPU and GPU (graphics processing unit) based systems.

    We find that GPUs can yield a significant speed-up (up to a factor of $10$ in the dense case) for these types of problems. In the sparse case GPUs are only advantageous for large problem sizes and the achieved speed-ups are more modest. In most cases the commutator-free variant is superior but especially on the GPU this advantage is rather small. In fact, none of the advantage of commutator-free methods on GPUs (and on multi-core CPUs) is due to the elimination of commutators. This has important consequences for the design of more efficient numerical methods.

\end{abstract}

\section{Introduction}

To numerically solve the Schr\"{o}dinger equation with a time-dependent Hamiltonian 
\begin{align}\label{eq:schroedinger}
\rmi \frac{\rmd\psi}{\rmd t}= H(t) \psi(t), \quad \psi(0)=\psi_0, 
\end{align} 
is a problem of significant interest in various fields of quantum mechanics. Applications range from discrete spin systems to (continuous) models of atom-laser interaction. Therefore, it is important to have both, good numerical algorithms, as well as an efficient implementation on state of the art computer hardware of these algorithms at one's disposal.

Magnus integrators are used in many such applications (see, for example, \cite{raedt2008,BCOR09,kormann2008,laptyeva2016,hl03}). The implementation of these Magnus integrators (which constitute a subclass of exponential integrators; for more details see \cite{ho10,HLW06,BCOR09}) requires the computation of the action of matrix exponentials in an efficient and stable manner. \textcolor{black}{For some problems, e.g. if the continuous Schr\"odinger equation is used to model atom-laser interaction, this can be done using fast Fourier techniques. However, for many other interesting problems this is not possible. For the latter case a} number of approaches have been proposed in the literature (see, for example, \cite{ho10,higham2008,hochbruck1997,al2011,caliari2014,farquhar2016}). Most of them are based on polynomial interpolation. In \cite{caliari2014,ckor15} it was shown that interpolation at Leja points is a very efficient way of performing this approximation for the Schr\"odinger equation. This algorithm interpolates the exponential function and thus reduces the task of computing the action of a matrix exponential to the task of computing a sequence of matrix-vector products. Let us also note that, in addition to the Schr\"{o}dinger equation considered in this paper, Magnus integrators have been successfully applied to many related problems. 

In addition to the matrix exponential, traditional Magnus integrators of higher order require the computation of matrix commutators (see, for example, \cite{ho10,blanes2000,BCOR09}). In time dependent problems (as those considered here) these matrix commutators have to be computed once every time step. Thus, especially for large problem sizes the corresponding cost can outweigh the cost of the matrix-vector products. Recently, commutator-free Magnus integrators have been developed \cite{bm06,alvermann2011}. They eliminate commutators but usually require additional matrix-vector products.

Due to the trend towards CPUs with more and more cores as well as the trend towards GPUs, providing an efficient implementation of numerical algorithms on modern multi-core CPUs and GPUs is of great practical importance. Some preliminary work on implementing exponential integrators \cite{einkemmer2013} and matrix functions \cite{farquhar2016} has been conducted on GPUs (with generally promising results). The purpose of the present work is to investigate the performance of both commutator-free and traditional Magnus integrators. This is done in the context of multi-core CPUs and GPUs. Although from a computational complexity point of view, one might conjecture that computing the commutators will dominate the total computational cost, this is not necessarily true in an actual implementation. In particular, on GPUs matrix-matrix products (necessary for computing the commutators) can operate close to peak efficiency while this is usually not the case for matrix-vector products. The comparison will be performed in the context of both short-range interactions (which lead to sparse Hamiltonians $H(t)$) and long-range interactions (which lead to dense Hamiltonians $H(t)$) to ascertain in which situations GPUs result in a significant gain in performance.

This paper is based, in part, on the thesis \cite{Auer16} and is structured as follows. In section \ref{sec:magnus} we provide an introduction to Magnus integrators and specify the numerical methods used in the subsequent sections. Section \ref{sec:implementation} then details the numerical approximation and the implementation. The numerical results are presented and discussed in section \ref{sec:ne}. Finally, we conclude in section \ref{sec:conclusion}.

\section{Magnus integrators\label{sec:magnus}}
The solution of the linear differential equation 
\begin{align}\label{eq:dgl}
Y'(t) = A(t)Y(t), \quad Y(0)=Y_0,
\end{align}
can be expressed as
\begin{align}\label{eq:gen_sol}
Y(t) = \exp \left(\Om(t)\right) Y_0,
\end{align}
where the difficulty lies in finding a suitable matrix $\Om(t)$.
In \cite{m54} Magnus used the ansatz of differentiating \eqref{eq:gen_sol} to 
find an expression for $\Om(t)$.
This results in 
\begin{align}\label{eq:Om_sol}
Y'(t) = \frac{\rmd}{\rmd t} \exp(\Om(t))Y_0 
	= \dexp_{\Om(t)}(\Om'(t)) \exp(\Om(t))Y_0,
\end{align}
where the operator $\dexp$ is defined as 
\begin{align}\label{eq:dexp}
\dexp_\Om(C) = \sum_{k=0}^\infty \frac{1}{(k+1)!} \ \ad_\Om^k (C) 
	= \varphi_1(\ad_\Om)(C),
\end{align} 
see \cite{HLW06}.
Here the operator $\ad_\Om^k(C)$ is the iterated 
commutator and recursively defined as
\begin{align*}
\ad^j_\Om(C)=\left[\Om, \ad_\Om^{j-1}(C)\right],\quad j\geq1,
\end{align*}
and $\ad^0_\Om(C)=C$.
Comparing \eqref{eq:dgl} and \eqref{eq:Om_sol} leads to 
\begin{align}
A(t) = \dexp_{\Om(t)}(\Om'(t)), \quad \Om(0)=0. 
\end{align}
By applying the inverse of the derivative of the matrix exponential we obtain 
a differential equation for $\Om$.
In fact, when $\|\Om(t)\|<\pi$ the operator $\dexp_{\Om(t)}$ is invertible and
has the convergent series representation 
\begin{align*}
\dexp_{\Om(t)}^{-1} (A(t)) = 
	\sum_{k=0}^\infty \frac{\beta_k}{k!} \ad_{\Om(t)}^k (A(t)),
\end{align*}
where $\beta_k$ denote the Bernoulli numbers. 
As a result we obtain an explicit differential equation for $\Om(t)$ as 
\begin{align}\label{eq:Om_dgl}
\begin{aligned}
\Om'(t) &= \dexp_{\Om(t)}^{-1} (A(t))\\ 
		&= A(t) -\frac{1}{2} \left[ \Om(t), A(t)\right] + 
			\frac{1}{12} \left[ \Om(t), \left[ \Om(t),A(t)\right] \right] + 
			\cdots\ .
\end{aligned}
\end{align}
Equation \eqref{eq:Om_dgl} can be integrated by Picard iteration and this leads 
to the so-called \emph{Magnus expansion},
\begin{align}\label{eq:magnusexpansion}
\begin{aligned}
\Om(t) = \int_0^t A(t_1) \rmd t_1 
	&- \frac{1}{2} \int_0^t \left[ \int_0^{t_1} A(t_2) \rmd t_2 , A(t_1) \right] \rmd t_1 \\
&+ \frac{1}{4} \int_0^t \left[ \int_0^{t_1} \left[ \int_0^{t_2} A(t_3)\rmd t_3, A(t_2)\right] \rmd t_2 , A(t_1) \right] \rmd t_1 \\
&+ \frac{1}{12} \int_0^t \left[ \int_0^{t_1}A(t_2)\rmd t_2,\left[ \int_0^{t_1} A(t_2)\rmd t_2, A(t_1)\right] \right] \rmd t_1 + \cdots\ .
\end{aligned}
\end{align}
To derive numerical methods from the Magnus expansion we assume a constant 
time step size $\tau$ and thus the solution after one time step is
\begin{align}\label{eq:onetimestep}
Y(t_n+\tau) = \exp(\Om(t_n+\tau)) Y(t_n),
\end{align}
resulting in the numerical scheme 
\begin{align}\label{eq:Om_n}
Y_{n+1}=\exp(\Om_n)Y_n,
\end{align} 
for a suitable approximation $\Om_n$ of $\Om(t_n+\tau)$.
One way of deriving a formula for $\Om_n$ is to approximate the integrals in 
\eqref{eq:magnusexpansion} by quadrature rules. 

In the following we will introduce the three traditional Magnus integrators that are used for the
numerical experiments in Section~\ref{sec:ne}.

\begin{met}[M2]\label{eg:M2}
The first example is the simplest method, which is obtained by 
truncating the series \eqref{eq:magnusexpansion} after the first term
and approximating the integral by the midpoint rule. This yields 
\begin{align*}
\Om_n(\tau) = \tau A\left( t_n+\frac{\tau}{2}\right) 
\end{align*}
as an approximation of $\Om(t_n + \tau)$.
The corresponding numerical scheme is the exponential midpoint rule
\begin{align*}
Y_{n+1} = \exp\left( \tau A\left( t_n+\frac{\tau}{2}\right) \right)  Y_n, 
\end{align*}
which is of order two.
\end{met}

\begin{met}[M4]\label{eg:M4}
The second example is a scheme of order four. 
The Magnus series \eqref{eq:magnusexpansion} is truncated after the second term 
and the integrals are approximated by the two-stage Gauss quadrature rule 
with weights $b_1=b_2=\frac{1}{2}$ 
and nodes $c_1=\frac{1}{2}-\frac{\sqrt{3}}{6}$,
$c_2= \frac{1}{2}+\frac{\sqrt{3}}{6}$.
We obtain
\begin{align*}
Y_{n+1} = \exp \left( \frac{\tau}{2} \left( A_1+A_2\right) + \frac{\sqrt{3}\tau^2}{12} \left[A_2,A_1\right] \right) Y_n,
\end{align*}
where $A_1 = A(t_n+c_1\tau)$ and $A_2 = A(t_n+c_2\tau)$.
\end{met}

\begin{met}[M6]\label{eg:M6}
As a third example, we consider the following scheme of order six:
\begin{align*}
\begin{split}
Y_{n+1} = &\exp \Bigg( B_1 + \frac{1}{2} B_3 \\ 
& \qquad + \frac{1}{240} \bigg[ -20 B_1 - B_3 + [ B_1, B_2 ], B_2 - \frac{1}{60} \Big[ B_1, 2 B_3 + [ B_1, B_2] \Big] \bigg] \Bigg) Y_n,
\end{split}
\end{align*}
where $A_i$ is an approximation of $A\left(t_n+c_i\tau\right)$ for 
\begin{align*}
c_1 = \frac{1}{2}-\frac{\sqrt{15}}{10},\quad
c_2 = \frac{1}{2},\quad
c_3 = \frac{1}{2}+\frac{\sqrt{15}}{10},
\end{align*}
and
\begin{align*}
B_1 &= \tau A_2,\quad
B_2 = \frac{\sqrt{15}}{3} \tau (A_3 -A_1),\quad
B_3 = \frac{10}{3} \tau (A_3 -2 A_2 + A_1).
\end{align*}
\end{met}

	As has been discussed in the introduction, the (nested) commutators arising in methods \ref{eg:M4} and \ref{eg:M6} can be expensive to compute. Thus, in addition, we consider commutator free Magnus methods (see \cite{bm06}). To achieve this our aim is to find an approximation of the form
\begin{align*}
Y(t_n + \tau) \approx Y_{n+1} = 
	\exp\Big(\Om_n^{(1)}\Big) \exp\Big(\Om_n^{(2)}\Big) \cdots 
	\exp\Big(\Om_n^{(s)}\Big) Y_n,
\end{align*}
with 
\begin{align*}
\Om_n^{(i)} = \tau \sum_{j=1}^J \alpha_{ij} A_j, 
\quad \text{ for } \quad i=1,\ldots,s.
\end{align*}
Here $A_i= A(t_n + c_i \tau)$ again denotes the evaluation of the matrix 
$A$ at a certain time. 
More precisely, the goal is to find $\Om_n^{(1)}, \ldots, \Om_n^{(s)}$ such that
\begin{align*}
\exp\Big(\Om_n^{(1)}\Big) \exp\Big(\Om_n^{(2)}\Big) \cdots \exp\Big(\Om_n^{(s)}\Big) 
= \exp(\Om_n) + \mathcal{O}\big(\tau^{S+1}\big)
\end{align*}
by determining $s$, $J$ and $\alpha_{ij}$ for $i = 1,\ldots,s$ 
and $j = 1, \ldots, J$. 
The matrix $\Om_n \approx \Om(t_n + \tau)$ is the matrix at time step $n$ 
of an $S$th order Magnus method given as in \eqref{eq:Om_n}. \textcolor{black}{In the numerical experiments we will use the following two numerical methods that have been derived in \cite{bm06} and \cite{alvermann2011}.}

\begin{met}[Cf4]\label{eq:cf4}
For $c_1=\frac{1}{2}-\frac{\sqrt{3}}{6}$, $c_2= \frac{1}{2}+\frac{\sqrt{3}}{6}$,
as well as $\alpha_1 = \frac{3-2 \sqrt{3}}{12}$,  and 
$\quad \alpha_2 = \frac{3+2 \sqrt{3}}{12}$ a commutator-free method of order 4 
is given by 
\begin{align*}
Y_{n+1} &= \exp\Big(\Om_n^{(1)}\Big) \exp\Big(\Om_n^{(2)}\Big) Y_n \\
	&= \exp\Big(\tau(\alpha_1 A_1 + \alpha_2 A_2)\Big) \exp\Big(\tau ( \alpha_2 A_1 + \alpha_1 A_2)\Big) Y_n.
\end{align*}
\end{met}

{\color{black}
\begin{met}[Cf4:3]\label{eq:cf4:3}
\textcolor{black}{This is an optimized commutator-free method of order 4 that uses three exponentials (see \cite[Sect. 5.2]{alvermann2011}).}
Let $A_i$ be an approximation of $A\left(t_n+c_i\tau\right)$ for 
\begin{align*}
c_1 = \frac{1}{2}-\frac{\sqrt{15}}{10},\quad
c_2 = \frac{1}{2},\quad
c_3 = \frac{1}{2}+\frac{\sqrt{15}}{10},
\end{align*}
and 
\begin{align*}
\alpha =
\begin{bmatrix} 
\left(\frac{37}{240}-\frac{10\sqrt{15}}{261}\right) & -\frac{1}{30} & 
    \left(\frac{37}{240}+\frac{10\sqrt{15}}{261}\right) \\
-\frac{11}{360} & \frac{23}{45} & -\frac{11}{360} \\
\left(\frac{37}{240}+\frac{10\sqrt{15}}{261}\right) & -\frac{1}{30} & 
    \left(\frac{37}{240}-\frac{10\sqrt{15}}{261}\right)
\end{bmatrix}.
\end{align*}
Then, the Cf4:3 method is given by
\begin{align*}
Y_{n+1} =& \exp\Big(\Om_n^{(1)}\Big) \exp\Big(\Om_n^{(2)}\Big) 
\exp\Big(\Om_n^{(3)}\Big) Y_n \\
    =& \exp\Big(\tau(\alpha_{11} A_1 + \alpha_{12} A_2 + \alpha_{13} A_3)\Big) 
    \exp\Big(\tau(\alpha_{21} A_1 + \alpha_{22} A_2 + \alpha_{23} A_3)\Big)\\
    &
    \exp\Big(\tau(\alpha_{31} A_1 + \alpha_{32} A_2 + \alpha_{33} A_3)\Big) Y_n.
\end{align*}
\end{met}
}
\begin{figure}[htb]
\centering
\pgfplotsset{height=0.4\linewidth,width=0.96\linewidth,compat=1.10,
    every axis/.append style={legend style={
        /tikz/every even column/.append style={column sep=9pt}}}}
\tikzsetnextfilename{order}
\begin{tikzpicture}[scale=1]%
    \begin{loglogaxis}[legend style={at={(1.00,1.17)}}, legend columns=0,cycle list name=short_list, xmin=0.004, xmax=0.25,  grid=major,ymin=1e-15, ymax=0.000013, xlabel={time step size $\tau$}, ylabel={err}]
      \addplot table[header=false, row sep=\\]{5.000000000e-01 2.220505e-05\\ 
          2.500000000e-01 3.777462e-06\\ 1.250000000e-01 9.029367e-07\\ 
          6.250000000e-02 2.237774e-07\\ 3.125000000e-02 5.582812e-08\\ 
          1.562500000e-02 1.394985e-08\\ 7.812500000e-03 3.487015e-09\\ 
          3.906250000e-03 8.717246e-10\\ 1.953125000e-03 2.179282e-10\\ 
          9.765625000e-04 5.448078e-11\\ 4.882812500e-04 1.361906e-11\\ 
          2.441406250e-04 3.403697e-12\\ 1.220703125e-04 8.501133e-13\\ };
        \addlegendentry{M2}
        \addplot table[header=false, row sep=\\]{5.000000000e-01 1.34120e-05\\ 
          2.500000000e-01 3.68260e-07\\ 1.250000000e-01 1.75768e-08\\ 
          6.250000000e-02 1.05899e-09\\ 3.125000000e-02 6.56255e-11\\ 
          1.562500000e-02 4.09545e-12\\ 7.812500000e-03 2.59227e-13\\ 
          3.906250000e-03 2.91663e-14\\ 1.953125000e-03 2.27846e-14\\ 
          9.765625000e-04 2.25747e-14\\ 4.882812500e-04 2.21701e-14\\ 
          2.441406250e-04 2.19862e-14\\ 1.220703125e-04 1.29114e-14\\ };
          \addlegendentry{M4}
        \addplot table[header=false, row sep=\\]{5.000000000e-01 1.34323e-05\\ 
          2.500000000e-01 1.56040e-07\\ 1.250000000e-01 1.08973e-09\\ 
          6.250000000e-02 1.58063e-11\\ 3.125000000e-02 2.44760e-13\\ 
          1.562500000e-02 2.31038e-14\\ 7.812500000e-03 2.26508e-14\\ 
          3.906250000e-03 2.26469e-14\\ 1.953125000e-03 2.26491e-14\\ 
          9.765625000e-04 2.25614e-14\\ 4.882812500e-04 2.21697e-14\\ 
          2.441406250e-04 2.19862e-14\\ 1.220703125e-04 1.29114e-14\\ };
          \addlegendentry{M6}
        \addplot table[header=false, row sep=\\]{5.000000000e-01 4.56729e-06\\ 
          2.500000000e-01 1.03562e-07\\ 1.250000000e-01 4.26112e-09\\ 
          6.250000000e-02 2.50504e-10\\ 3.125000000e-02 1.54315e-11\\ 
          1.562500000e-02 9.61446e-13\\ 7.812500000e-03 6.42878e-14\\ 
          3.906250000e-03 2.29833e-14\\ 1.953125000e-03 2.25699e-14\\ 
          9.765625000e-04 2.21698e-14\\ 4.882812500e-04 2.19866e-14\\ 
          2.441406250e-04 1.29116e-14\\ 1.220703125e-04 3.60148e-15\\ };
        \addlegendentry{Cf4} 
        \addplot table[header=false, row sep=\\]{0.50000000  5.02824e-07\\ 
          0.25000000  5.70012e-09\\ 0.12500000  9.07108e-11\\ 
          0.06250000  4.59001e-12\\ 0.03125000  2.68682e-13\\ 
          0.01562500  2.43487e-14\\ 0.00781250  2.23865e-14\\ 
          0.00390625  2.26192e-14\\ 0.00195312  2.23435e-14\\ 
          0.00097656  2.20575e-14\\ };   
          \addlegendentry{Cf4:3}           
        \addplot coordinates{(1e-4,1e-11) (1, 0.001)};
          \addlegendentry{$\tau^2$}
        \addplot coordinates{(1e-3,1e-15) (1, 0.001)};
      \addlegendentry{$\tau^4$}
      \addplot coordinates{(1e-3,1e-23) (1, 0.001)};
          \addlegendentry{$\tau^6$}
    \end{loglogaxis}
\end{tikzpicture}
\caption{\label{fig:orderplot_dense}
Order plot for the methods M2 (blue with dots), M4 (green with squares), 
M6 (orange with triangles), Cf4 (brown with diamonds), and CF4:3 (red with 
star). 
For Example~\ref{eg:dense} with $N=2^{10}$ and 
varying time step size $\tau$.}
\end{figure}
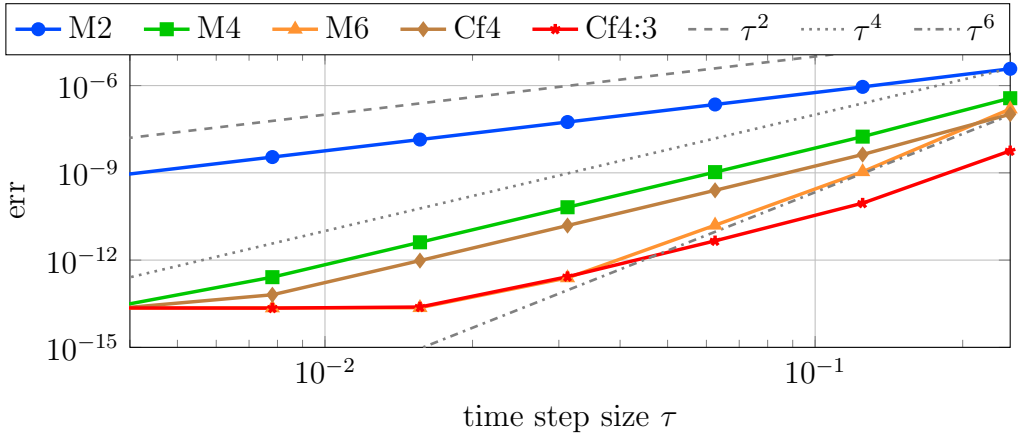

In Figure~\ref{fig:orderplot_dense} we show the error as a function of the time step size. This serves to check the order of the \textcolor{black}{five} methods discussed in this section and helps in validating our implementation. We also note that the Cf4 method requires two matrix exponentials but no commutator and is somewhat more accurate compared to the M4 method (which requires only a single matrix exponential but the computation of one-commutator).
{\color{black} Furthermore, we can see that Cf4:3, which requires three matrix exponentials, has the same order as the other fourth order methods but a smaller error constant. In fact, in this experiment the error is even below or on the same level as the order six method. }

\section{Numerical approximation and implementation \label{sec:implementation}}

So far we have not discussed how to approximate the action of the matrix exponential. That is, how to compute
\begin{align*}
Y_{n+1}=\exp(\Om_n)Y_n.
\end{align*}
It is not viable to compute $\exp(\Om_n)$ and then apply it to the vector $Y_n$. Furthermore, if $\Om_n$ is sparse, this is, in general, not the case for $\exp(\Om_n)$ and thus a significant penalty in terms of memory is implied. However, a number of approaches to efficiently compute the action of a matrix exponential have been developed in the literature. Probably the most prominent are Krylov subspace methods, interpolation at Chebyshev points, the truncated Taylor method, and interpolation at Leja points. For all of these methods the main cost to approximate $Y_{n+1}$ are matrix-vector products.

In the numerical results presented in the next section we will exclusively use interpolation at Leja point (the method described in \cite{ckor15}). This method is favorably compared to interpolation at Chebyshev points as the order of the interpolation polynomial can be chosen adaptively. In addition, it has been shown in \cite{caliari2014} that interpolation at Leja points can be superior to using Krylov methods. In addition, in contrast to Krylov subspace methods, only few vectors have to be kept in memory. This reduces the memory footprint of the Leja interpolation which is especially helpful for the GPU implementation, where memory is a more scarce resource compared to traditional CPU systems.

The main idea of the Leja method is to approximate the exponential by an 
interpolation polynomial. The interpolation nodes $\xi_0,\ldots,\xi_m$ for a polynomial of degree $m$ are chosen to be Leja points (see \cite{ckor15}, for a precise definition). The corresponding interpolation polynomial is expressed in Newton form and therefore the main cost of the algorithm is contained in a short term recurrence formula that performs matrix-vector products. 

Let $p_m(A)v$ be the interpolation polynomial of degree $m$ that approximates the action of the exponential of the matrix $A$ applied to the vector $v$. Furthermore, we denote by $d_i:=\exp[\xi_0, \xi_1, \ldots, \xi_i]$ the divided differences of order $i$. Then Leja interpolation has the form 
\begin{align*}
	\text{exp}(A)v \approx p_m(A)v= \sum_{i=0}^m d_i\prod_{j=0}^{i-1}(A-\xi_jI)v,
\end{align*}
and can be implemented by the following code fragment
{\color{black}
\begin{algorithmic}[1]
\Require{$v$, $A$}
\Ensure{$w$}
\State{$w=v$}
\State{$p=d_0v$}
\For{$i=1\colon m$}
    \State{$w=A w-\xi_{i-1}w$}
    \State{$p=p+d_i w$}
\EndFor
\end{algorithmic}
The control structures are executed in sequence (on the CPU), while the vector and matrix operations (lines 1, 2, 4, and 5) are parallelized.}
The actual implementation is more involved. It makes sure that the backward error of the interpolation is below a user specified tolerance (up to the unit round-off). Furthermore, an early termination criterion is implemented to reduce the cost further. For more details we refer the reader to \cite{ckor15}.

It should be noted that the Leja method is not entirely matrix-free. In fact the matrix itself is required for estimating the necessary parameters of the interpolation (i.e.~an estimate on the spectrum of the matrix). Nevertheless, we can think of these parameters as fixed and therefore the performance of the method breaks down to performing the above code fragment as fast as possible. 

The speed of the numerical method is highly dependent on the underlying implementation of matrix-vector products (to compute the Leja interpolation) and matrix-matrix products (to compute the commutators). The Leja interpolation routines are implemented as part of the \emph{expleja} project\footnote{see \url{https://bitbucket.org/expleja/expleja}} in a generic fashion. That is, for all of the implementations the core Leja algorithm is the same. We use the preprocessor to switch out
the calls to the matrix-vector and matrix-matrix product routines. Thus, the library is designed in such a way that it is easy to include your favorite library dealing with matrices, vectors, and most importantly the matrix-vector and matrix-matrix products. 

In this paper we are concerned with traditional CPUs and GPUs as well as dense and sparse matrices. In the numerical experiments in the next section, we use the standard BLAS\footnote{Version: gcc 4.8.4, libblas-dev: 1.2.20110419-7} and Intel Math Kernel Library (MKL)\footnote{Version: icc 15.0.0 20140723, MKL 2017.0.098} \cite{MKL} routines for dense matrices on the CPU, SuiteSparse\footnote{Version: gcc 4.8.4, libsuitesparse-dev: 1:4.2.1-3ubuntu1} \cite{d06} and Intel MKL for spare matrices on the CPU, and the CUDA libraries cuBLAS and cuSPARSE\footnote{Version: nvcc 7.0.27, CUDA Toolkit 7.5} \cite{cuBlas,cuSparse} for dense and sparse matrices, respectively, on the GPU.

The probably most widely used sparse matrix storage scheme is the compressed sparse row (CSR) format (see, for example, \cite{cuSparse}). This format can store arbitrary sparse matrices and is extensively used in a range of applications. In addition, the CSR format enjoys universal support from almost all sparse matrix libraries.
It should be noted, however, that specific sparse matrix formats have been developed that allow for a more favorable memory access pattern and consequently improved performance on GPUs. However, as recent research \cite{greathouse2014,guo2016,zhang2016} shows the potential improvement compared to a well implemented CSR algorithm is at best 60 \%. In addition, the magnitude of improvement is highly problem dependent. Due to this and due to ubiquity of the CSR format we will make exclusive use of it in the present paper.

To allow a fair comparison and to enable other scientists to take advantage of our implementation the code is available via the \emph{expleja} project.%

Let us further note that in the case of the GPU implementation we make sure that we move data from the CPU to the GPU (or in the other direction) only if necessary. More specifically, we transfer all input data at the beginning of the computation, perform all the numerical computation on the GPU, and finally only the output data is transferred back to the CPU. Consequently, in all instances the data transfer from and to the GPU takes only a negligible amount of time.

The following hardware is used in all of the experiments\\[5pt]
\fbox{\begin{minipage}{.5\linewidth}
	CPU
	\begin{itemize}
        \item 2$\times$ Intel Xeon E5-2630 CPU 
		\item 2$\times$ 8 cores
		\item 32GB Ram
	\end{itemize}
\end{minipage}
\begin{minipage}{.5\linewidth}
	GPU
	\begin{itemize}
		\item Tesla K80
		\item 4992 CUDA Cores
		\item 2$\times$ 12 GB RAM
	\end{itemize}
\end{minipage}}\\[5pt]

Before proceeding, it is instructive to discuss the performance characteristics of the two main ingredients in the Magnus integrator. Namely, the matrix-vector products required for the Leja interpolation and the matrix-matrix products required for assembling the matrix (the latter are only required for the Magnus integrators that involve commutators). Since the matrix is time dependent we have to assemble the matrix (and thus compute the commutators) in every time step. A straightforward complexity analysis leads us to conclude that the matrix-matrix products dominate the run time of the algorithm. In fact, this is consistent with what we will observe in the next section for the sequential version of the code. Since for traditional higher order Magnus integrators the number of required commutators can be quite large, research has been conducted in both reducing \cite{blanes2000} and completely eliminating (commutator-free approach) \cite{bm06} commutators from the numerical method.

However, this analysis is too simplistic to accurately reflect the performance of the numerical method in an actual implementation. The first point to make here is that potentially a large number of matrix-vector products are needed in each time step. This is in particular true for matrices with large eigenvalues in the left half of the complex plane (i.e.~for stiff problems). This behavior alone is sufficient to make a purely theoretical analysis difficult. However, there is a second aspect that needs to be taken into account on modern computer hardware (both on multi-core CPUs and on GPUs). Namely, that the matrix-matrix products constitute a compute bound problem and thus can be parallelized very efficiently, while matrix-vector products are memory bound and are thus limited by the (much slower) memory bandwidth. 

To be more specific let us consider the dense case. We have implemented a test case for both the matrix-vector as well as matrix-matrix products. The parallel implementation using Intel MKL on the CPU saturates with only 4 cores (although 16 are available on above mentioned system; a behavior that is characteristic for memory bound problems) achieving approximately 10 GFlop/s. This constitutes approximately 70 \% of the theoretical memory bandwidth but only a small fraction of the GFlop/s that are available on that system. On the other hand, the parallel implementation of the matrix-matrix products using Intel MKL yields approximately $150$ GFlop/s.
Thus, while the computational complexity of the matrix-matrix products is far worse, in a parallel setting, the constants are potentially much smaller. To conclude this section let us note that essentially the same behavior occurs on the GPU and that we will revisit this issue again in the next section when we discuss the numerical results.

\section{Numerical experiments \label{sec:ne}}

In the following discussion ``host'' refers to the CPU and 
``device'' refers to the GPU. 
For the Intel MKL implementation on the CPU we tested various number of threads 
and used the configuration that gives the best performance. 
The time of the computation is measured by taking the difference of calling 
\texttt{clock()} before and after the time integration.
The result is converted to seconds. 
For all of the experiments we compute a reference solution on the GPU with the 
method M6 and a fixed time step size of $\tau=10^{-5}$.
If not specified explicitly, the Leja method is used with a relative and absolute 
tolerance of $10^{-10}$ and the early termination criterion is activated. 

In order to illustrate the behavior of the considered Magnus integrators and
their corresponding implementation we consider the \textcolor{black}{discrete} Schr\"odinger equation
\begin{align*}
\mathrm{i} \partial_t \psi(t) = H(t) \psi(t), \quad \psi(0)=\psi_0
\end{align*} 
\textcolor{black}{with the Hamiltonian in the form $H(t)=H_1+h(t)H_2$}. More specifically, we consider a Heisenberg model. This model is often used in quantum mechanics to describe the spins of a 
magnetic system. A spin describes a magnetic dipole, where the direction of the spin corresponds
to the magnetic moment.

We investigate the performance for both sparse and dense matrices $H(t)$. Physically, this corresponds to a short-range (sparse) and long-range (dense) interaction. We will refer to these situations as the local and non-local model, respectively.
As each spin is described by a two dimensional (complex) vector the resulting 
Hamiltonian $H(t)$ is described by a $2^n\times 2^n$ complex matrix,
where $n$ corresponds to the amount of spins in the model. 
In the following, we use the notation $N=2^n$.

We discuss the particular shape of the two models used in the corresponding examples. 
Both rely on the three Pauli matrices 
\begin{align}\label{eq:Pauli}
\sigma^x = 
\begin{pmatrix}
0 & 1\\
1 & 0\\
\end{pmatrix}, \quad 
\sigma^y = 
\begin{pmatrix}
0 & -i\\
i & 0\\
\end{pmatrix} \quad \text{and} \quad
\sigma^z = 
\begin{pmatrix}
1 & 0\\
0 & -1
\end{pmatrix}.
\end{align}
Furthermore we define
\begin{align}\label{eq:sigma^alpha}
\sigma_j^{\alpha} = 
(\otimes_{i=1}^{j-1} I) \otimes \sigma^{\alpha} \otimes (\otimes_{i=j+1}^{n} I), 
\end{align}
for $\alpha \in \{x,y,z\}$, where $\otimes$ denotes the tensor product.
As initial value $\psi_0 \in \C^{2^n}$, we use 
\begin{align}\label{eq:initial_value}
\psi_0 = 
\begin{pmatrix}
1&0,&0&1,& \psi_0^5& \psi_0^6, &\psi_0^7 & \ldots
\end{pmatrix}^\mathrm{T},
\end{align}
where the entries $\psi_0^5$ to $\psi_0^N$ are chosen randomly such that 
$\left\|\begin{pmatrix}
\psi_0^i & 
\psi_0^{i+1}
\end{pmatrix}\right\|_2 = 1$ for $i$ odd.
The initial value is chosen such that the spins $1$ and $2$ 
are in a defined state (up and down, respectively).
Due to the coupling with the other spins these two spins will leave this pure 
state and move into a neutral position. 
This behavior can be observed in the numerical experiments.

In all of our experiments we split up the matrix $H(t)$ as $H(t)=H_1+h(t)H_2$ 
where the matrices $H_1$ and $H_2$ are computed before the actual 
time integration on either the host or the device. 
\textcolor{black}{This form of the Hamiltonian means that the $H(t)$, for a specific $t$, can be assembled relatively quickly (compared to the computational cost of performing the Leja interpolation). We expect that the performance results presented here generalize to other problems where this is the case. Note that for the GPU implementation}
the matrix $H(t)$ is formed on the device without 
transferring any data from the host. 
Furthermore, the commutators required for the various tested Magnus integrators 
are computed directly on the device as well. 

{\color{black}
As an example, the following pseudocode illustrates the implementation of one time step of the M4 scheme
\begin{algorithmic}[1]
\Require{$Y_n, H_1, H_2$, $t_n$, $\tau$}
\Ensure{$Y_{n+1}$}

\State{$d_1$ = $h(t_n + c_1 \tau)$}
\State{$d_2$ = $h(t_n + c_2 \tau)$}
\State{$A_1$ = $H_1$ + $d_1 H_2$}
\State{$A_2$ = $H_1$ + $d_2 H_2$}
\State{$A_3$ = $\tfrac{1}{2} \tau (A_1 + A_2)$}
\State{$A_4$ = $A_2 A_1$}
\State{$A_5$ = $A_1 A_2$}
\State{$A_6$ = $\tfrac{\sqrt{3}}{12} \tau^2 (A_4-A_5)$}
\State{$A_7$ = $A_3$ + $A_6$}
\State{$Y_{n+1}$ = Leja($A_7$,$Y_n$)}
\end{algorithmic}
where the function \texttt{Leja} approximates the action of the matrix exponential (the pseudocode is given in section \ref{sec:implementation}). Lines 1 and 2 (which are computational extremely cheap) are computed sequentially (on the CPU), while the remainder of the algorithm consists of matrix operations that are all executed in parallel (either on the CPU or on the GPU). Note that the variables $A_1$ to $A_7$ are introduced in the above code fragment for clarity. In the actual implementation variables are reused whenever possible in order to decrease the amount of storage required.
}

\begin{bsp}[Dense matrix case]\label{eg:dense}
For the dense case, we use a non-local Heisenberg model, 
where all spins are coupled. 
This results in a dense matrix. 
With the help of the Pauli matrices and $\sigma_j^{\alpha}$, see \eqref{eq:Pauli} 
and \eqref{eq:sigma^alpha}, we state the non-local Heisenberg model
\begin{align*}
H(t) = 
- \sum_{i=1}^n \underset{j \neq i}{\sum_{j=1}^n} J_{ij} \sigma_i^z \sigma_j^z 
- \sum_{i=1}^n h(t) \sigma_i^x \quad \in\C^{2^n \times 2^n},
\end{align*}
    where $J_{ij} = \frac{1}{|i-j|}$ and $h(t) = \sin(\omega t)$ \textcolor{black}{with $\omega=1$}.
The initial value $\psi_0 \in \C^{2^n}$ is given by \eqref{eq:initial_value}.
We note that the matrix $H(t)$ consists of two time independent parts that are 
coupled by $h(t)$.
In particular we have $H(t)=H_1 + h(t) H_2$, for 
\begin{align*}
H_1 = 
- \sum_{i=1}^n \underset{j \neq i}{\sum_{j=1}^n} J_{ij} \sigma_i^z \sigma_j^z
 \quad\text{and}\quad
H_2 = - \sum_{i=1}^n \sigma_i^x.
\end{align*}
We exploit this structure in our implementation by assembling the matrices 
$H_1$ and $H_2$ before the actual time integration. 
This allows us to considerably reduce the computational cost. 

In the experiments illustrated in Figures~\ref{fig:dense_cvdf_parallel} 
and~\ref{fig:dense_cvdf_methods} we fix the time step size to $\tau=10^{-3}$ 
and vary the size of the matrices from $N=2^6$ to $N=2^{13}$.

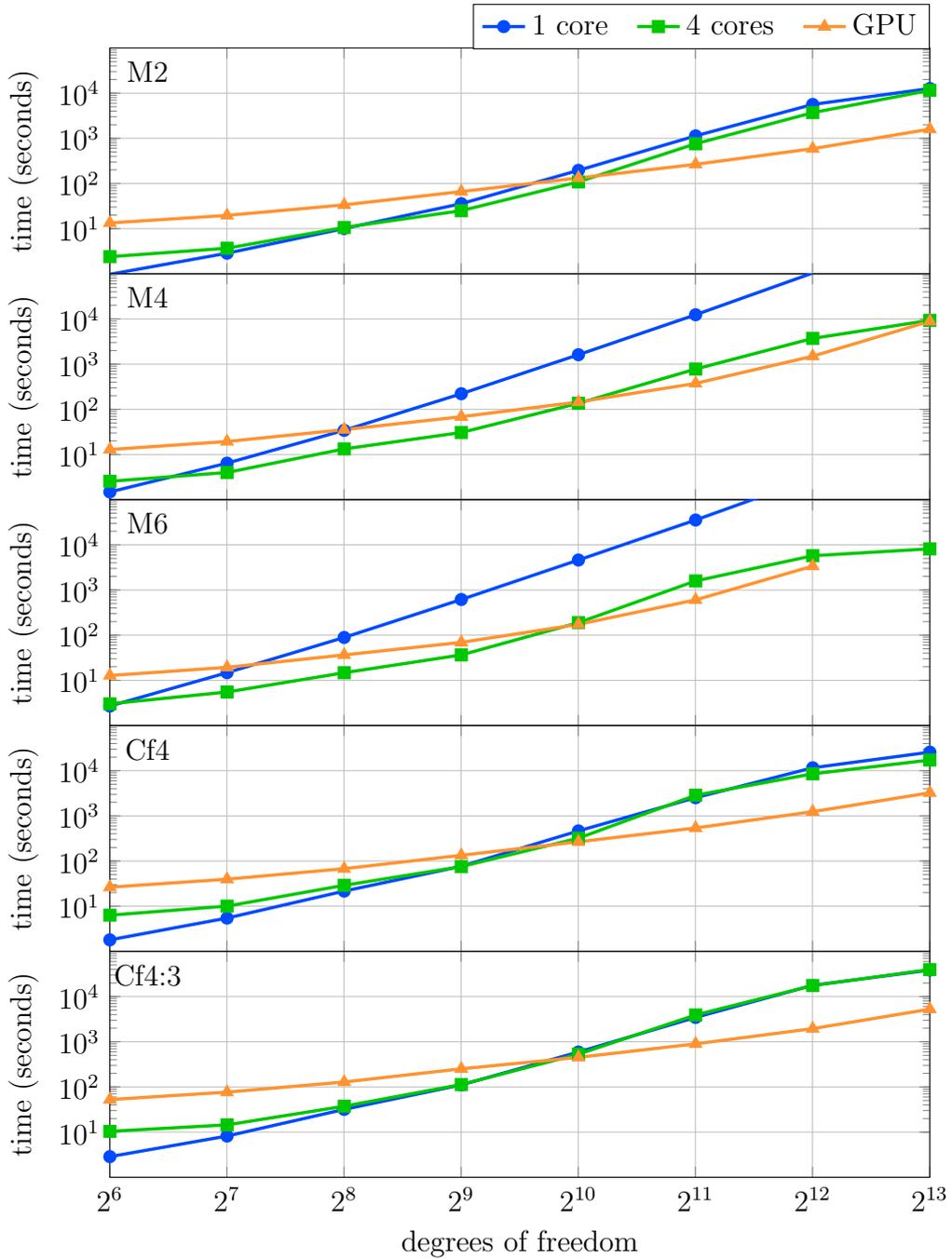
\begin{figure}[htb]
\pgfplotsset{height=0.35\linewidth,width=0.96\linewidth,compat=1.10,
    every axis/.append style={legend style={
        /tikz/every even column/.append style={column sep=9pt}}}}
\tikzsetnextfilename{df_vs_t_parallel_dense}
\begin{tikzpicture}[scale=1]
\begin{groupplot}[group style={
                      group name=myplot,
                      group size= 1 by 5,
                      vertical sep=0cm}]
    \nextgroupplot[ymode=log, xmode=log, cycle list name=short_list,
      log basis x={2},
      legend style={at={(1.00,1.195)}}, legend columns=0,
      xmin=64, xmax=8192, grid=major, ymin=1, ymax=1e+5, 
      ytick={10,100,1000,10000},
      xticklabel=\empty, ylabel={time (seconds)}
      ]
      \node[] at (axis cs:80,3e+4){M2};
        \addplot table[header=false, row sep=\\]{64  0.975606\\128  2.856344\\
          256  10.045418\\512  35.675247\\1024  195.296173\\2048  1133.559570\\
          4096  5623.276367\\8192 12676.941406\\};
          \addlegendentry{1 core}
        \addplot table[header=false, row sep=\\]{64  2.393702\\128  3.680825\\
          256  10.604962\\512  25.012636\\1024  108.552162\\2048  758.061951\\
          4096  3702.694092\\8192 11539.749023\\};
          \addlegendentry{4 cores}
        \addplot table[header=false, row sep=\\]{64  13.361777\\128  19.645292\\
          256  33.580009\\512  66.238617\\1024  132.351440\\2048  265.393097\\
          4096  589.680176\\8192 1595.803223\\};
          \addlegendentry{GPU}
        \pgfplotsset{cycle list shift=1}
    \nextgroupplot[ymode=log, xmode=log, cycle list name=short_list,
      log basis x={2},
      xmin=64, xmax=8192, grid=major, ymin=1, ymax=1e+5, 
      ytick={10,100,1000,10000},
      xticklabel=\empty, ylabel={time (seconds)}
      ]
      \node[] at (axis cs:80,3e+4){M4};
        \addplot table[header=false, row sep=\\]{64  1.480639\\128  6.437018\\
          256  34.292519\\512  221.503067\\1024  1617.999634\\
          2048  12400.294922\\4096  103984.015625\\};
        \addplot table[header=false, row sep=\\]{64  2.555689\\128  3.999207\\
          256  13.293523\\512  30.728449\\1024  136.186081\\2048 779.189819\\
          4096 3715.239014\\8192 9333.510742\\};
        \addplot table[header=false, row sep=\\]{64  12.888658\\128  19.413528\\
          256  35.543831\\512  68.585022\\1024  145.340195\\2048  375.094116\\
          4096  1500.037476\\8192 8894.645508\\};
        \pgfplotsset{cycle list shift=1}
    \nextgroupplot[ymode=log, xmode=log, cycle list name=short_list,
      log basis x={2},
      xmin=64, xmax=8192, grid=major, ymin=1, ymax=1e+5, 
      ytick={10,100,1000,10000},
      xticklabel=\empty, ylabel={time (seconds)}
      ]
      \node[] at (axis cs:80,3e+4){M6};
        \addplot table[header=false, row sep=\\]{64  2.694856\\128  14.771204\\
          256  88.957367\\512  615.639099\\1024  4612.224609\\
          2048  35461.652344\\ 4096 302420.031250\\ 8192 2436301.25\\};
        \addplot table[header=false, row sep=\\]{64  3.023151\\128  5.510114\\
          256  14.731993\\512  36.514553\\1024  189.262070\\2048  1586.070068\\
          4096 5736.422363\\8192 8153.743164\\};
        \addplot table[header=false, row sep=\\]{64  12.772597\\128  19.408417\\
          256  36.547245\\512  69.262718\\1024  173.413239\\2048  604.220825\\
          4096 3385.124756\\};
        \pgfplotsset{cycle list shift=1}
    \nextgroupplot[ymode=log, xmode=log, cycle list name=short_list,
      log basis x={2},
      xmin=64, xmax=8192, grid=major, ymin=1, ymax=1e+5, 
      ytick={10,100,1000,10000},
      ylabel={time (seconds)},
      xticklabel=\empty
      ]
      \node[] at (axis cs:80,3e+4){Cf4};
        \addplot table[header=false, row sep=\\]{64  1.797878\\128  5.453630\\
          256  21.589592\\512  76.428169\\1024  464.804169\\2048  2524.949951\\
          4096 11604.466797\\8192 25753.390625\\};
        \addplot table[header=false, row sep=\\]{64  6.315959\\128  10.039067\\
          256  28.990112\\512  75.394493\\1024  320.491272\\2048 2864.920166\\
          4096 8522.568359\\8192 17225.064453\\};
        \addplot table[header=false, row sep=\\]{64  26.336920\\128  39.609924\\
          256  67.716011\\512  134.079269\\1024  266.515564\\2048  539.278564\\
          4096  1237.559570\\8192 3245.654053\\};
        \pgfplotsset{cycle list shift=1}
    \nextgroupplot[ymode=log, xmode=log, cycle list name=short_list,
      log basis x={2},
      xmin=64, xmax=8192, grid=major, ymin=1, ymax=1e+5, 
      ytick={10,100,1000,10000},
      ylabel={time (seconds)},xlabel={degrees of freedom}
      ]
      \node[] at (axis cs:80,3e+4){Cf4:3};
        \addplot table[header=false, row sep=\\]{64  2.8631931\\128  8.180897\\
          256 31.854588\\512 111.019287\\ 1024 593.168701\\ 2048 3454.531738\\
          4096  17744.652344\\8192  38535.203125\\};
        \addplot table[header=false, row sep=\\]{64 10.405783\\ 128   14.470025\\ 
          256   37.431385\\ 512   113.033592\\ 1024  530.420898\\ 
          2048  3920.310547\\ 4096  17531.533203\\ 8192  39669.656250\\};
        \addplot table[header=false, row sep=\\]{64  52.496048\\128  76.746605\\
          256  128.608459\\512  251.819092\\1024  451.819092\\2048  899.317505\\
          4096  1952.542114\\8192 5271.891602\\};
        \pgfplotsset{cycle list shift=1}
      
\end{groupplot}
\end{tikzpicture}
\caption{\label{fig:dense_cvdf_parallel}%
Computational cost vs.\ degrees of freedom for three different 
parallelizations of the five integrators M2, M4, M6, Cf4, and CF4:3, 
grouping by the different methods.
The data correspond to Example~\ref{eg:dense} (dense matrix) with fixed 
time step size $\tau=10^{-3}$ for computing $\psi(1)$.}
\end{figure}
In the first experiment we take a look how the parallelization influences the 
method. 
\textcolor{black}{In Figure~\ref{fig:dense_cvdf_parallel} we can observe that for large enough 
    matrices all methods benefit from GPU acceleration. Where exactly this benefit kicks in depends on the method. One would assume that the speed-up is most pronounced for methods M4 and M6 that need the computation of matrix commutators. However, since these schemes also profit the most from CPU parallelization, the relative gain of the GPU implementation is actually smaller compared to M2, Cf4, and Cf4:3.}

\begin{figure}[htb]
\pgfplotsset{height=0.4\linewidth,width=0.96\linewidth,compat=1.10,
    every axis/.append style={legend style={
        /tikz/every even column/.append style={column sep=9pt}}}}
\tikzsetnextfilename{df_vs_t_methods_dense}
\begin{tikzpicture}[scale=1]
\begin{groupplot}[group style={
                      group name=myplot,
                      group size= 1 by 4,
                      vertical sep=0cm}]
    \nextgroupplot[ymode=log, xmode=log, cycle list name=short_list,
      log basis x={2},
      legend style={at={(1.00,1.17)}}, legend columns=0,
      xmin=64, xmax=8192, grid=major, ymin=1, ymax=1e+5, 
      ytick={10,100,1000,10000},
      xticklabel=\empty, ylabel={time (seconds)}
      ]
      \node[] at (axis cs:100,3e+4){1 core};
        \addplot table[header=false, row sep=\\]{64  0.975606\\128  2.856344\\
          256  10.045418\\512  35.675247\\1024  195.296173\\2048  1133.559570\\
          4096  5623.276367\\8192 12676.941406\\};
          \addlegendentry{M2}
        \addplot table[header=false, row sep=\\]{64  1.480639\\128  6.437018\\
          256  34.292519\\512  221.503067\\1024  1617.999634\\
          2048  12400.294922\\4096  103984.015625\\};
          \addlegendentry{M4}
        \addplot table[header=false, row sep=\\]{64  2.694856\\128  14.771204\\
          256  88.957367\\512  615.639099\\1024  4612.224609\\2048  35461.652344\\
          4096 302420.031250\\ 8192 2436301.25\\};
          \addlegendentry{M6}
        \addplot table[header=false, row sep=\\]{64  1.797878\\128  5.453630\\
          256  21.589592\\512  76.428169\\1024  464.804169\\2048  2524.949951\\
          4096 11604.466797\\ 8192 25753.390625\\};
          \addlegendentry{Cf4}
        \addplot table[header=false, row sep=\\]{64  2.8631931\\128  8.180897\\
          256 31.854588\\512 111.019287\\ 1024 593.168701\\ 2048 3454.531738\\
          4096  17744.652344\\8192  38535.203125\\};
          \addlegendentry{Cf4:3}
    \nextgroupplot[ymode=log, xmode=log, cycle list name=short_list,
      log basis x={2},
      xmin=64, xmax=8192, grid=major, ymin=1, ymax=1e+5, 
      ytick={10,100,1000,10000},
      xticklabel=\empty, ylabel={time (seconds)}
      ]
      \node[] at (axis cs:100,3e+4){4 cores};
        \addplot table[header=false, row sep=\\]{64  2.393702\\128  3.680825\\
        256  10.604962\\512  25.012636\\1024  108.552162\\2048  758.061951\\
        4096  3702.694092\\8192 11539.749023\\};
        \addplot table[header=false, row sep=\\]{64  2.555689\\128  3.999207\\
        256  13.293523\\512  30.728449\\1024  136.186081\\2048 779.189819\\
        4096 3715.239014\\8192 9333.510742\\};
        \addplot table[header=false, row sep=\\]{64  3.023151\\128  5.510114\\
        256  14.731993\\512  36.514553\\1024  189.262070\\2048  1586.070068\\
        4096 5736.422363\\8192 8153.743164\\};
        \addplot table[header=false, row sep=\\]{64  6.315959\\128  10.039067\\
        256  28.990112\\512  75.394493\\1024  320.491272\\2048 2864.920166\\
        4096 8522.568359\\8192 17225.064453\\};
        \addplot table[header=false, row sep=\\]{64 10.405783\\ 128   14.470025\\ 
          256   37.431385\\ 512   113.033592\\ 1024  530.420898\\ 
          2048  3920.310547\\ 4096  17531.533203\\ 8192  39669.656250\\};
    \nextgroupplot[ymode=log, xmode=log, cycle list name=short_list,
      log basis x={2},
      xmin=64, xmax=8192, grid=major, ymin=1, ymax=1e+5, 
      ytick={10,100,1000,10000},
      xlabel={degrees of freedom}, ylabel={time (seconds)}
      ]
      \node[] at (axis cs:100,3e+4){GPU};
        \addplot table[header=false, row sep=\\]{64  13.361777\\128  19.645292\\
        256  33.580009\\512  66.238617\\1024  132.351440\\2048  265.393097\\
        4096  589.680176\\8192 1595.803223\\};
        \addplot table[header=false, row sep=\\]{64  12.888658\\128  19.413528\\
        256  35.543831\\512  68.585022\\1024  145.340195\\2048  375.094116\\
        4096  1500.037476\\8192 8894.645508\\};
        \addplot table[header=false, row sep=\\]{64  12.772597\\128  19.408417\\
        256  36.547245\\512  69.262718\\1024  173.413239\\2048  604.220825\\
        4096 3385.124756\\};
        \addplot table[header=false, row sep=\\]{64  26.336920\\128  39.609924\\
        256  67.716011\\512  134.079269\\1024  266.515564\\2048  539.278564\\
        4096  1237.559570\\8192 3245.654053\\};
        \addplot table[header=false, row sep=\\]{64  52.496048\\128  76.746605\\
          256  128.608459\\512  251.819092\\1024  451.819092\\2048  899.317505\\
          4096  1952.542114\\8192 5271.891602\\};
        \addplot table[header=false, row sep=\\]{64  52.496048\\128  76.746605\\
          256  128.608459\\512  251.819092\\1024  451.819092\\2048  899.317505\\
          4096  1952.542114\\8192 5271.891602\\};
\end{groupplot}
\end{tikzpicture}
\caption{\label{fig:dense_cvdf_methods}%
Computational cost vs.\ degrees of freedom for three different 
parallelizations of the five integrators M2, M4, M6, CF4, and Cf4:3,
grouping by different parallelizations. 
The data correspond to Example~\ref{eg:dense} with fixed 
time step size $\tau=10^{-3}$ for computing $\psi(1)$.}
\end{figure}
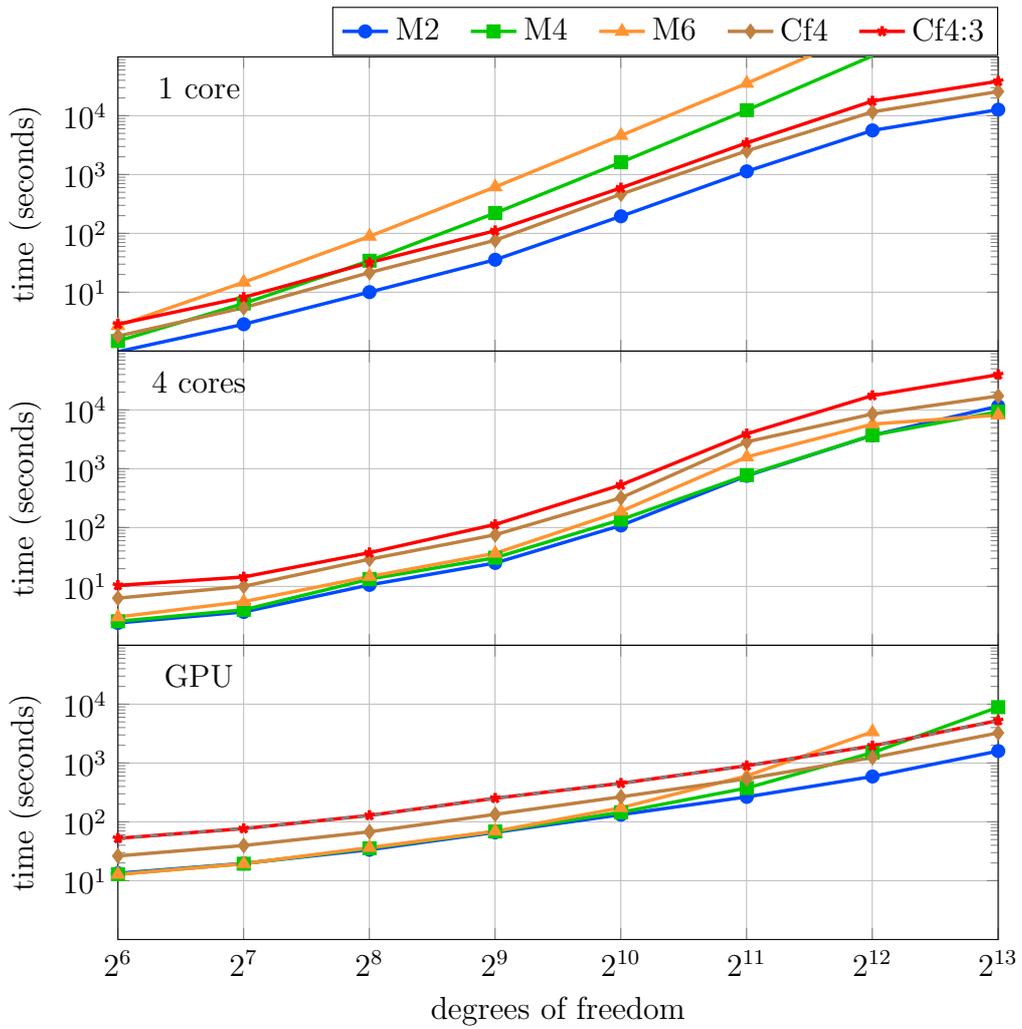
In Figure~\ref{fig:dense_cvdf_methods} we show how the five different methods 
compare on the same hardware. 
If we first focus on the single core implementation (top) we can see that 
M2 has the least computational cost. 
In fact the cost grows as $\mathcal{O}(N^2)$ which roughly corresponds 
to the cost of the matrix-vector products in the Leja interpolation.
This is also the case for the \textcolor{black}{Cf4/Cf4:3 methods}, where no commutator is required but 
two\textcolor{black}{/three} matrix exponentials have to be computed. 
As a result the method needs more time per time step compared to M2 but has approximately the same growth rate. 
The methods M4 and M6 grow with order three, corresponding to the cost of 
the matrix-matrix products arising in the commutators.

By using 4 cores (middle) we can see that the cost of the methods move 
together. M2 and M4 need about the same, M6 slightly more, \textcolor{black}{Cf4 is even more expensive, and Cf4:3 turns out to be the most expensive method (per time step).} 
In general, we observe a growth rate of about $\mathcal{O}(N^2)$ for all methods
as the matrix-matrix products can be parallelized very efficiently. 

If we move to the GPU (bottom) we observe that the growth for all methods is 
linear until about $2^{10}$ where full capacity of the GPU is reached. 
After this we observe that the different methods start to behave similar as for
the single CPU, but they are much faster.

\begin{figure}[htb]
\pgfplotsset{height=0.7\linewidth,width=0.96\linewidth,compat=1.10,
    every axis/.append style={legend style={
        /tikz/every even column/.append style={column sep=9pt}}}}
\tikzsetnextfilename{tol_vs_cpu_dense}
\begin{tikzpicture}[scale=1]%
    \begin{customlegend}[legend style={at={(11.58,8.7)}},
    legend columns=0, 
    legend style={/tikz/every even column/.append style={column sep=9pt}}, 
    legend entries={1 core, 4 cores, GPU, M2, M4, M6, Cf4, Cf4:3}]
       \addlegendimage{black,line width=1pt, solid, sharp plot}
       \addlegendimage{black,line width=1pt, dashed, sharp plot}
       \addlegendimage{black,line width=1pt, dotted, sharp plot}
       \addlegendimage{mblue,line width=1pt, mark=*, only marks}
       \addlegendimage{mgreen,line width=1pt, mark=square*, only marks}
       \addlegendimage{morange,line width=1pt, mark=triangle*, only marks}
       \addlegendimage{brown,line width=1pt, mark=diamond*, only marks}
       \addlegendimage{mred,line width=1pt, mark=star, only marks}
   \end{customlegend}
    \begin{loglogaxis}[legend style={at={(1.00,1.17)}}, 
    legend columns=0,cycle list name=long_list2, 
    xmin=8e-14, xmax=5e-4,  grid=major,ymin=6e-1, 
    xlabel={accuracy}, ylabel={time (seconds)},
    mark options={solid,scale=0.7}]
      \addplot table[header=false, row sep=\\]{6.8176e-05  1.1637e+01\\
        7.4599e-06  1.9965e+01\\9.0293e-07  5.1587e+01\\9.9303e-08  1.5247e+02\\
        9.8919e-09  4.5158e+02\\9.9183e-10  1.4021e+03\\9.9955e-11  4.3718e+03\\
        9.9998e-12  1.3280e+04\\9.9982e-13  4.1280e+04\\9.9996e-14 1.3720e+05\\};
      \addplot table[header=false, row sep=\\]{6.1933e-05 1.1607e+02\\
        2.0335e-06 3.2390e+02\\3.6826e-07 4.3061e+02\\5.8025e-08 6.4428e+02\\
        7.0701e-09 1.0682e+03\\8.2985e-10 1.8163e+03\\9.7351e-11 3.1000e+03\\
        9.3973e-12 5.5269e+03\\9.6099e-13 9.7656e+03\\9.7792e-14 1.7184e+04\\};  
      \addplot table[header=false, row sep=\\]{1.4877e-04 3.1283e+02\\
        1.3432e-05 6.1531e+02\\
        1.4178e-06 9.1920e+02\\1.5604e-07 1.2086e+03\\3.0139e-08 1.5115e+03\\
        7.0600e-09 1.8114e+03\\5.2578e-10 2.7155e+03\\9.0422e-11 3.6166e+03\\
        7.7602e-12 5.4242e+03\\8.4766e-13 7.9432e+03\\9.0214e-14 1.1427e+04\\};
      \addplot table[header=false, row sep=\\]{2.5881e-05 1.9074e+01\\
        2.5881e-05 1.9527e+01\\2.5881e-05 2.0041e+01\\2.5881e-05 2.0576e+01\\
        4.5672e-06 4.4940e+01\\6.6416e-07 6.2326e+01\\3.3358e-08 9.1513e+01\\
        7.4684e-09 1.4382e+02\\8.0396e-10 2.3500e+02\\8.3730e-11 3.8948e+02\\
        9.6242e-12 6.5182e+02\\9.6144e-13 1.1701e+03\\9.8188e-14 2.0585e+03\\};
    \addplot table[header=false, row sep=\\]{9.6286e-06 2.1542e+01\\ 
        9.6286e-06 2.3581e+01\\ 9.6286e-06 2.1779e+01\\ 9.6286e-06 2.4200e+01\\ 
        9.6286e-06 2.5123e+01\\ 5.0282e-07 4.8088e+01\\ 5.2362e-08 6.0083e+01\\ 
        5.7001e-09 8.4731e+01\\ 3.7276e-10 1.2241e+02\\ 9.0710e-11 1.6037e+02\\ 
        7.9939e-12 2.8253e+02\\ 8.7033e-13 4.6087e+02\\ 9.7407e-14 8.0225e+02\\}; 
      \addplot table[header=false, row sep=\\]{6.8176e-05 6.7318e+00\\
        7.4599e-06 1.2503e+01\\9.0293e-07 3.5160e+01\\9.9303e-08 1.0129e+02\\
        9.8919e-09 3.2862e+02\\9.9183e-10 1.0494e+03\\9.9955e-11 3.2463e+03\\
        9.9998e-12 1.0425e+04\\9.9982e-13 2.9002e+04\\9.9996e-14 8.7305e+04\\};
      \addplot table[header=false, row sep=\\]{6.1933e-05 4.2012e+01\\
        2.0335e-06 1.1037e+02\\3.6826e-07 1.4717e+02\\5.8025e-08 2.2061e+02\\
        7.0701e-09 3.6363e+02\\8.2985e-10 6.1224e+02\\9.7351e-11 1.0493e+03\\
        9.3973e-12 1.8747e+03\\9.6099e-13 3.3015e+03\\9.7792e-14 5.8268e+03\\};  
      \addplot table[header=false, row sep=\\]{1.4877e-04 1.0509e+02\\
        1.3432e-05 1.9493e+02\\
        1.4178e-06 2.9459e+02\\1.5604e-07 3.9606e+02\\3.0139e-08 4.9491e+02\\
        7.0600e-09 5.9198e+02\\5.2578e-10 8.9046e+02\\9.0422e-11 1.1875e+03\\
        7.7602e-12 1.7738e+03\\8.4766e-13 2.5574e+03\\9.0214e-14 3.7759e+03\\};
      \addplot table[header=false, row sep=\\]{2.5881e-05 1.1325e+01\\
        4.5672e-06 2.5522e+01\\6.6416e-07 3.5849e+01\\3.3358e-08 5.3800e+01\\
        7.4684e-09 7.9586e+01\\8.0396e-10 1.3741e+02\\8.3730e-11 2.6840e+02\\
        9.6242e-12 4.8332e+02\\9.6144e-13 6.9588e+02\\9.8188e-14 1.2297e+03\\};
      \addplot table[header=false, row sep=\\]{9.6286e-06 2.2436e+01\\ 
        9.6286e-06 2.1440e+01\\ 9.6286e-06 1.9754e+01\\ 9.6286e-06 1.9416e+01\\ 
        9.6286e-06 2.2995e+01\\ 5.0282e-07 4.0881e+01\\ 5.2362e-08 5.3427e+01\\ 
        5.7001e-09 7.4170e+01\\ 3.7276e-10 1.2442e+02\\ 9.0710e-11 1.1607e+02\\ 
        7.9939e-12 2.1868e+02\\ 8.7033e-13 3.4663e+02\\ 9.7407e-14 6.0265e+02\\};
      \addplot table[header=false, row sep=\\]{6.8176e-05 7.3125e-01\\
        7.4599e-06 1.7870e+00\\ 9.0293e-07 4.6749e+00\\ 9.9303e-08 1.4552e+01\\
        9.8919e-09 4.4745e+01\\ 9.9183e-10 1.4173e+02\\ 9.9955e-11 4.4017e+02\\
        9.9998e-12 1.3749e+03\\ 9.9982e-13 4.4871e+03\\ 9.9996e-14 1.3973e+04\\};
      \addplot table[header=false, row sep=\\]{6.1933e-05 1.6469e+00\\
        2.0335e-06 4.4732e+00\\ 3.6826e-07 5.9183e+00\\ 5.8025e-08 9.2682e+00\\
        7.0701e-09 1.5299e+01\\ 8.2985e-10 2.5933e+01\\ 9.7351e-11 4.3994e+01\\
        9.3973e-12 7.8370e+01\\ 9.6099e-13 1.3980e+02\\ 9.7792e-14 2.5051e+02\\};  
      \addplot table[header=false, row sep=\\]{1.4877e-04 3.8246e+00\\
        1.3432e-05 6.7664e+00\\
        1.4178e-06 1.0096e+01\\ 1.5604e-07 1.3537e+01\\ 3.0139e-08 1.6971e+01\\
        7.0600e-09 2.0465e+01\\ 5.2578e-10 3.0490e+01\\ 9.0422e-11 4.0659e+01\\
        7.7602e-12 6.0833e+01\\ 8.4766e-13 8.7788e+01\\ 9.0214e-14 1.2772e+02\\};
      \addplot table[header=false, row sep=\\]{2.5881e-05 1.1708e+00\\
        4.5672e-06 2.7008e+00\\ 6.6416e-07 3.8245e+00\\ 3.3358e-08 5.8137e+00\\
        7.4684e-09 8.6946e+00\\ 8.0396e-10 1.5861e+01\\ 8.3730e-11 2.6869e+01\\
        9.6242e-12 4.6233e+01\\ 9.6144e-13 8.2160e+01\\ 9.8188e-14 1.4142e+02\\};
      \addplot table[header=false, row sep=\\]{9.6286e-06 2.5685e+00\\ 
        9.6286e-06 2.3314e+00\\ 9.6286e-06 2.2855e+00\\ 9.6286e-06 2.3106e+00\\ 
        9.6286e-06 2.3172e+00\\ 5.0282e-07 4.5758e+00\\ 5.2362e-08 5.6306e+00\\ 
        5.7001e-09 8.4186e+00\\ 3.7276e-10 1.1861e+01\\ 9.0710e-11 1.5648e+01\\ 
        7.9939e-12 2.8337e+01\\ 8.7033e-13 4.8187e+01\\ 9.7407e-14 8.1940e+01\\};
    \end{loglogaxis}
\end{tikzpicture}
\caption{\label{fig:dense_avsc}%
    Accuracy vs.\ computational cost for the five integrators M2, M4, 
M6, Cf4, and Cf4:3 when using different parallelization techniques. 
The data correspond to Example~\ref{eg:dense} with fixed 
matrix dimension $N=2^{12}$. For a fixed step size $\tau$ the error at $\psi(1)$ 
is measured. }
\end{figure}
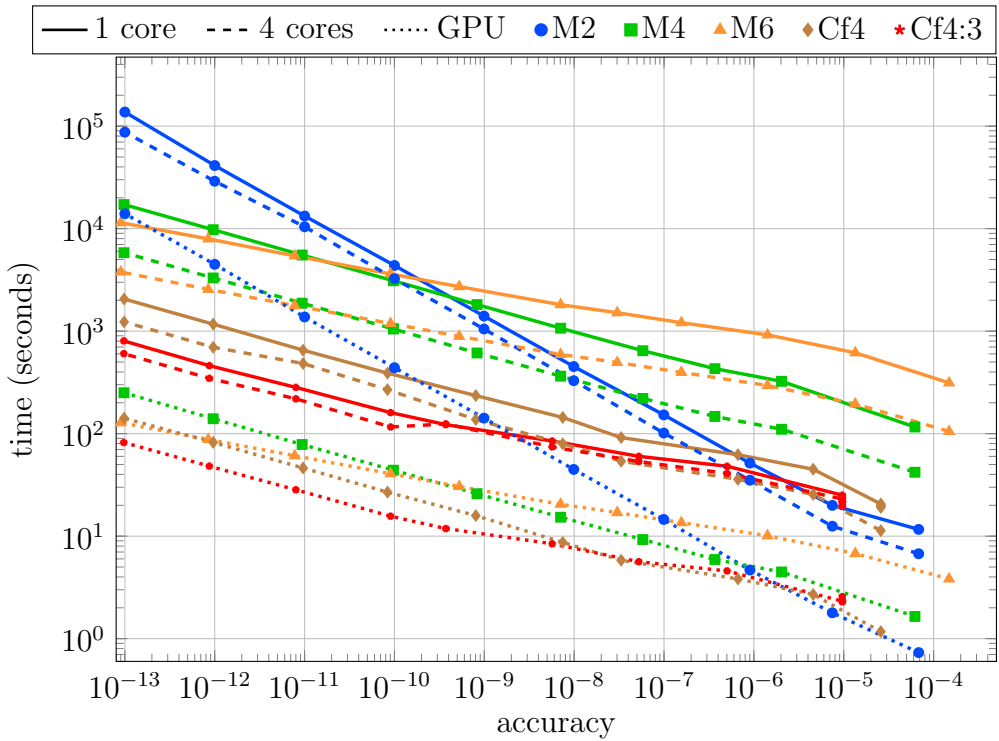
In our final experiment we change the setup.
This time we fix the matrix size to $N=2^{12}$, 
prescribe a certain tolerance, and compare the cost of each method, 
for each of the parallelization techniques,
to achieve the specified tolerance. 
The matrix size is chosen such that a GPU parallelization makes sense. 
We note that in all cases the Leja method with the prescribed tolerance 
divided by $10$ is used (that is, a more accurate approximation is used for the polynomial interpolation compared to the time integrator).
The results can be found in Figure~\ref{fig:dense_avsc}. 
\textcolor{black}{In the plot the line style corresponds to the different parallelizations; full for single-core CPU, dashed for multi-core CPU, and dotted for the GPU implementation. The marker shape and the color correspond to the different methods used; blue circles for M2, green squares for M4, orange triangles for M6, brown diamonds for Cf4 and red stars for Cf4:3.}

From this plot we can see that the GPU implementation is highly beneficial. 
For all methods the GPU achieves a speed-up of roughly 10 compared to the multi-core, 
and about 20 compared to the single-core implementation. 
Furthermore, we can observe that overall the \textcolor{black}{Cf4:3} method is the most efficient 
method in terms of accuracy versus computation time \textcolor{black}{(although, for tolerances above $10^{-8}$ the performance of Cf4 is identical to the performance of Cf4:3)}.
On the other hand it is clear that for high accuracy the M2 method is not 
favorable. This is due to the fact that this method is only second order accurate.

\textcolor{black}{For the GPU implementation the methods M4, M6, Cf4, and Cf4:3 are separated by no more than a factor of 4.}
For the parallel CPU implementations this is different. 
Here \textcolor{black}{Cf4:3} is clearly the fastest and the difference to M4 and M6 is approximately a factor of 5 and 10 
for the multi- and single-core CPU implementation, respectively. 

An important point to make here is that the majority of the computational advantage of the Cf4 and \textcolor{black}{Cf4:3} methods for multi-core CPUs and GPUs is due to its increased accuracy (see Figure \ref{fig:orderplot_dense}) and not due to any advantage in cost per time step. In fact, Figure \ref{fig:dense_cvdf_methods} shows that almost all of the advantage in cost (which is clearly significant in the sequential case) is lost once we consider the multi-core CPU or GPU implementation.

\end{bsp}

\begin{bsp}[Sparse matrix case]\label{eg:sparse}
For the sparse case we use a local Heisenberg model.
Here only neighboring spins are coupled and therefore the corresponding matrix
is sparse. 
With the help of the Pauli matrices, see \eqref{eq:Pauli} 
and \eqref{eq:sigma^alpha}, we state the local Heisenberg model
\begin{align*}
H(t) = -\frac{1}{2} \sum_{j=1}^n (J_x \sigma_j^x \sigma_{j+1}^x + 
	J_y \sigma_j^y \sigma_{j+1}^y + J_z \sigma_j^z \sigma_{j+1}^z + 
	h(t) \sigma_j^z) \quad \in \C^{2^n \times 2^n},
\end{align*}
    where $J_x, J_y, J_z \in \C$ and $h(t) = \sin(\omega t)$ \textcolor{black}{with $\omega=1$}.
The initial value is given in \eqref{eq:initial_value}. 
For the numerical tests we chose $J_x = 1$, $J_y = 2$, and $J_z = 3$.

Again we note that the matrix $H(t)$ consists of two time independent parts 
that are coupled by $h(t)$. 
We split-up the equation in a similar fashion as for the dense 
model in order to save computational cost. 

In the experiments illustrated by Figures~\ref{fig:sparse_cvdf_parallel},~%
\ref{fig:sparse_cvdf_methods}, and~\ref{fig:sparse_densevssparse} 
we fix the time step size to $\tau=10^{-3}$ 
and vary the size of the matrices from $N=2^6$ to $N=2^{14}$. 

\begin{figure}[htb]
\pgfplotsset{height=0.35\linewidth,width=0.96\linewidth,compat=1.10,
    every axis/.append style={legend style={
        /tikz/every even column/.append style={column sep=9pt}}}}
\tikzsetnextfilename{df_vs_t_parallel_sparse}
\begin{tikzpicture}[scale=1]
\begin{groupplot}[group style={
                      group name=myplot,
                      group size= 1 by 5,
                      vertical sep=0cm}]
    \nextgroupplot[ymode=log, xmode=log, cycle list name=short_list,
      log basis x={2},
      legend style={at={(1.00,1.22)}}, legend columns=0,
      xmin=64, xmax=16384, grid=major, ymin=1, ymax=1e+5, 
      ytick={10,100,1000,10000},
      xticklabel=\empty, ylabel={time (seconds)}
      ]
      \node[] at (axis cs:80,3e+4){M2};
        \addplot table[header=false, row sep=\\]{64  0.733815\\128  1.363036\\
          256  2.621308\\512  8.298598\\1024  19.680628\\2048  74.794983\\
          4096 135.792450\\8192 888.653076\\16384 1768.667480\\};
          \addlegendentry{1 core}
        \addplot table[header=false, row sep=\\]{64 0.985729\\128 1.401922\\
          256 2.586694\\512 6.454594\\1024 16.377220\\2048 50.132160\\
          4096 135.426453\\8192 674.241821\\16384 1681.959839\\};
          \addlegendentry{2 cores}
        \addplot table[header=false, row sep=\\]{64  8.855888\\128  11.498544\\
          256  13.053856\\512  17.386793\\1024  34.066345\\2048  70.845787\\
          4096  157.077972\\8192 347.498688\\16384 1429.036987\\};
          \addlegendentry{GPU}
        \pgfplotsset{cycle list shift=2}
        \addplot coordinates{(64,1) (16384, 65536)}; \addlegendentry{$N^2$}
    \nextgroupplot[ymode=log, xmode=log, cycle list name=short_list,
      log basis x={2},
      xmin=64, xmax=16384, grid=major, ymin=1, ymax=1e+5, 
      ytick={10,100,1000,10000},
      xticklabel=\empty, ylabel={time (seconds)}
      ]
      \node[] at (axis cs:80,3e+4){M4};
        \addplot table[header=false, row sep=\\]{64  1.191079\\128  2.785111\\
          256  6.540586\\512  24.383690\\1024  57.021210\\2048  201.901382\\
          4096 450.884369\\8192 2204.968994\\16384 4395.212891\\};
        \addplot table[header=false, row sep=\\]{64 1.372011\\128 2.535856\\
          256 5.338428\\512 13.220507\\1024 32.191692\\2048 100.522980\\
          4096 248.119659\\8192 1051.603882\\16384 2912.950684\\};
        \addplot table[header=false, row sep=\\]{64  9.311101\\128  13.936012\\
          256  16.465527\\512  26.607653\\1024  45.978233\\2048  120.764725\\
          4096  254.435654\\8192 464.477661\\16384 1931.938965\\};
        \pgfplotsset{cycle list shift=1}
    \nextgroupplot[ymode=log, xmode=log, cycle list name=short_list,
      log basis x={2},
      xmin=64, xmax=16384, grid=major, ymin=1, ymax=1e+5, 
      ytick={10,100,1000,10000},
      xticklabel=\empty, ylabel={time (seconds)}
      ]
      \node[] at (axis cs:80,3e+4){M6};
        \addplot table[header=false, row sep=\\]{64  2.720267\\128  11.224236\\
          256  56.237579\\512  317.672089\\1024 1339.070801\\2048 8604.895508\\
          4096 11880.875000\\8192 51486.000000\\16384 136573.859375\\};
        \addplot table[header=false, row sep=\\]{64 2.629126\\128 7.793422\\
          256 26.957670\\512 113.473251\\1024 569.103149\\2048 3333.650146\\
          4096 10935.640625\\8192 41970.199219\\16384 150634.296875\\};
        \addplot table[header=false, row sep=\\]{64  12.725996\\128  18.603924\\
          256  24.206373\\512  46.387238\\1024  94.505707\\2048  282.435883\\
          4096 827.340149\\8192 2910.669678\\16384 9673.436523\\};
        \pgfplotsset{cycle list shift=1}
    \nextgroupplot[ymode=log, xmode=log, cycle list name=short_list,
      log basis x={2},
      xmin=64, xmax=16384, grid=major, ymin=1, ymax=1e+5, 
      ytick={10,100,1000,10000}, xticklabel=\empty,
      ylabel={time (seconds)},
      ]
      \node[] at (axis cs:80,3e+4){Cf4};
        \addplot table[header=false, row sep=\\]{64 1.328777\\128 2.456156\\
          256 4.413183\\512 14.401124\\1024 33.225277\\2048 129.445190\\
          4096 233.285065\\8192 1530.429565\\16384 3243.655273\\};
        \addplot table[header=false, row sep=\\]{64 2.033957\\128 2.788319\\
          256 5.385729\\512 12.810895\\1024 30.103764\\2048 100.900169\\ 
          4096 272.240875\\8192 1412.255127\\16384 3301.049316\\};
        \addplot table[header=false, row sep=\\]{64  19.937572\\128  25.231447\\
          256  28.331539\\512  38.146656\\1024  74.376343\\2048  157.967682\\
          4096  356.126892\\8192 755.808960\\16384 2958.770020\\};
        \pgfplotsset{cycle list shift=1}
    \nextgroupplot[ymode=log, xmode=log, cycle list name=short_list,
      log basis x={2},
      xmin=64, xmax=16384, grid=major, ymin=1, ymax=1e+5, 
      ytick={10,100,1000,10000},
      ylabel={time (seconds)},xlabel={degrees of freedom}
      ]
      \node[] at (axis cs:80,3e+4){Cf4:3};
        \addplot table[header=false, row sep=\\]{64 1.832307\\ 128 3.418502\\ 
          256 7.635237\\ 512 20.592896\\ 1024 50.174290\\ 2048 183.552856\\ 
          4096 342.464783\\ 8192 2307.829102\\ 16384 4862.610352\\ };
        \addplot table[header=false, row sep=\\]{64 2.444384\\ 128 3.569566\\ 
          256 6.735585\\ 512 15.506566\\ 1024 39.090477\\ 2048 127.852898\\ 
          4096 328.842316\\ 8192 1694.070923\\ 16384 4533.560547\\ };
        \addplot table[header=false, row sep=\\]{64 25.924303\\ 128 32.700203\\ 
          256 34.643745\\ 512 51.845512\\ 1024 88.208603\\ 2048 183.384430\\ 
          4096 432.193695\\ 8192 978.193787\\ 16384 3977.904785\\ };
        \pgfplotsset{cycle list shift=1}
\end{groupplot}
\end{tikzpicture}
\caption{\label{fig:sparse_cvdf_parallel}
Computational cost vs.\ degrees of freedom for three different 
parallelizations of the five integrators M2, M4, M6, Cf4, and Cf4:3, 
grouping different parallelizations.
The data correspond to Example~\ref{eg:sparse} with fixed 
time step size $\tau=10^{-3}$ for computing $\psi(1)$.}
\end{figure}
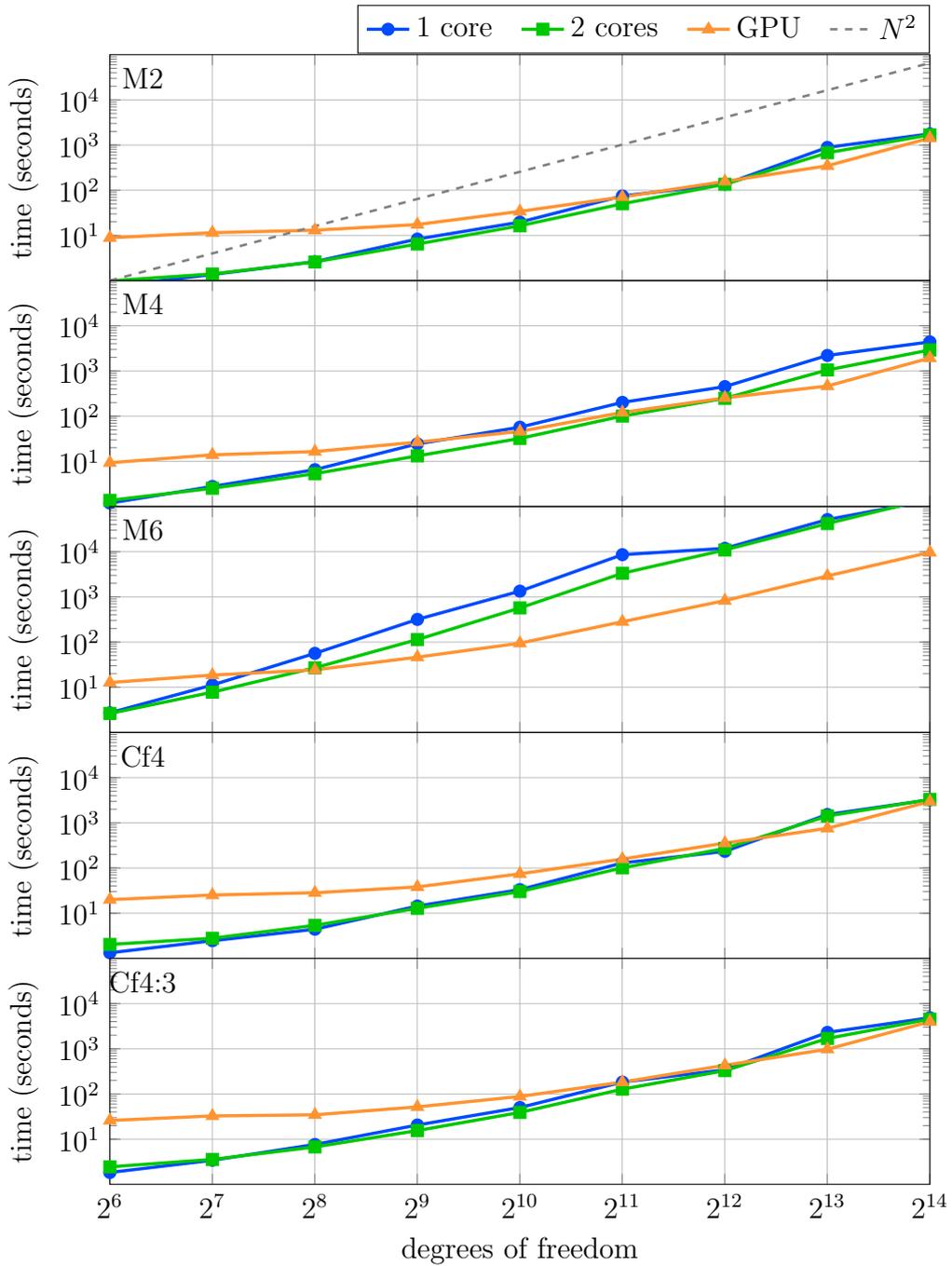
In the first experiment, illustrated in Figure~\ref{fig:sparse_cvdf_parallel}, 
we 
take a look on how the parallelization influences the performance of the methods. 
Here the GPU implementation does not achieve as high speed-ups as we observed for the dense case.
Only for the M6 method we can observe a significant gain for medium sized matrices. In this case we need to compute nested commutators for which the GPU acceleration is  highly beneficial. In all other cases the GPU is only favorable once we employ $2^{12}$ or more degrees of freedom. For small matrices the parallel CPU implementations are clearly faster compared to the GPU implementations. 

\begin{figure}[htb]
\pgfplotsset{height=0.4\linewidth,width=0.96\linewidth,compat=1.10,
    every axis/.append style={legend style={
        /tikz/every even column/.append style={column sep=9pt}}}}
\tikzsetnextfilename{df_vs_t_methods_sparse}
\begin{tikzpicture}[scale=1]
\begin{groupplot}[group style={
                      group name=myplot,
                      group size= 1 by 4,
                      vertical sep=0cm}]
    \nextgroupplot[ymode=log, xmode=log, cycle list name=short_list,
      log basis x={2},
      legend style={at={(1.00,1.17)}}, legend columns=0,
      xmin=64, xmax=16384, grid=major, ymin=1, ymax=1e+5, 
      ytick={10,100,1000,10000},
      xticklabel=\empty, ylabel={time (seconds)}
      ]
      \node[] at (axis cs:100,3e+4){1 core};
        \addplot table[header=false, row sep=\\]{64  0.733815\\128  1.363036\\
          256  2.621308\\512  8.298598\\1024  19.680628\\2048  74.794983\\
          4096 135.792450\\8192 888.653076\\16384 1768.667480\\};
          \addlegendentry{M2}
        \addplot table[header=false, row sep=\\]{64  1.191079\\128  2.785111\\
          256  6.540586\\512  24.383690\\1024  57.021210\\2048  201.901382\\
          4096 450.884369\\8192 2204.968994\\16384 4395.212891\\};
          \addlegendentry{M4}
        \addplot table[header=false, row sep=\\]{64  2.720267\\128  11.224236\\
          256  56.237579\\512  317.672089\\1024 1339.070801\\2048 8604.895508\\
          4096 11880.875000\\8192 51486.000000\\16384 136573.859375\\};
          \addlegendentry{M6}
        \addplot table[header=false, row sep=\\]{64 1.328777\\128 2.456156\\
          256 4.413183\\512 14.401124\\1024 33.225277\\2048 129.445190\\
          4096 233.285065\\8192 1530.429565\\16384 3243.655273\\};
          \addlegendentry{Cf4}
        \addplot table[header=false, row sep=\\]{64 1.832307\\ 128 3.418502\\ 
          256 7.635237\\ 512 20.592896\\ 1024 50.174290\\ 2048 183.552856\\ 
          4096 342.464783\\ 8192 2307.829102\\ 16384 4862.610352\\ };
         \addlegendentry{Cf4:3}
    \nextgroupplot[ymode=log, xmode=log, cycle list name=short_list,
      log basis x={2},
      xmin=64, xmax=16384, grid=major, ymin=1, ymax=1e+5, 
      ytick={10,100,1000,10000},
      xticklabel=\empty, ylabel={time (seconds)}
      ]
      \node[] at (axis cs:100,3e+4){2 cores};
        \addplot table[header=false, row sep=\\]{64 0.985729\\128 1.401922\\
          256 2.586694\\512 6.454594\\1024 16.377220\\2048 50.132160\\
          4096 135.426453\\8192 674.241821\\16384 1681.959839\\};
        \addplot table[header=false, row sep=\\]{64 1.372011\\128 2.535856\\
          256 5.338428\\512 13.220507\\1024 32.191692\\2048 100.522980\\
          4096 248.119659\\8192 1051.603882\\16384 2912.950684\\};
        \addplot table[header=false, row sep=\\]{64 2.629126\\128 7.793422\\
          256 26.957670\\512 113.473251\\1024 569.103149\\2048 3333.650146\\
          4096 10935.640625\\8192 41970.199219\\16384 150634.296875\\};
        \addplot table[header=false, row sep=\\]{64 2.033957\\128 2.788319\\
          256 5.385729\\512 12.810895\\1024 30.103764\\2048 100.900169\\
          4096 272.240875\\8192 1412.255127\\16384 3301.049316\\};
        \addplot table[header=false, row sep=\\]{64 2.444384\\ 128 3.569566\\ 
          256 6.735585\\ 512 15.506566\\ 1024 39.090477\\ 2048 127.852898\\ 
          4096 328.842316\\ 8192 1694.070923\\ 16384 4533.560547\\ };
    \nextgroupplot[ymode=log, xmode=log, cycle list name=short_list,
      log basis x={2},
      xmin=64, xmax=16384, grid=major, ymin=1, ymax=1e+5, 
      ytick={10,100,1000,10000},
      xlabel={degrees of freedom}, ylabel={time (seconds)}
      ]
      \node[] at (axis cs:100,3e+4){GPU};
        \addplot table[header=false, row sep=\\]{64  8.855888\\128  11.498544\\
          256  13.053856\\512  17.386793\\1024  34.066345\\2048  70.845787\\
          4096  157.077972\\8192 347.498688\\16384 1429.036987\\};
        \addplot table[header=false, row sep=\\]{64  9.311101\\128  13.936012\\
          256  16.465527\\512  26.607653\\1024  45.978233\\2048  120.764725\\
          4096  254.435654\\8192 464.477661\\16384 1931.938965\\};
        \addplot table[header=false, row sep=\\]{64  12.725996\\128  18.603924\\
          256  24.206373\\512  46.387238\\1024  94.505707\\2048  282.435883\\
          4096 827.340149\\8192 2910.669678\\16384 9673.436523\\};
        \addplot table[header=false, row sep=\\]{64  19.937572\\128  25.231447\\
          256  28.331539\\512  38.146656\\1024  74.376343\\2048  157.967682\\
          4096  356.126892\\8192 755.808960\\16384 2958.770020\\};
        \addplot table[header=false, row sep=\\]{64 25.924303\\ 128 32.700203\\ 
          256 34.643745\\ 512 51.845512\\ 1024 88.208603\\ 2048 183.384430\\ 
          4096 432.193695\\ 8192 978.193787\\ 16384 3977.904785\\ };
\end{groupplot}
\end{tikzpicture}
\caption{\label{fig:sparse_cvdf_methods}
Computational cost vs.\ degrees of freedom for three different 
parallelization of the five integrators M2, M4, M6, Cf4, and Cf4:3,
grouping different methods. 
The data correspond to Example~\ref{eg:sparse} with fixed 
time step size $\tau=10^{-3}$ for computing $\psi(1)$.}
\end{figure}
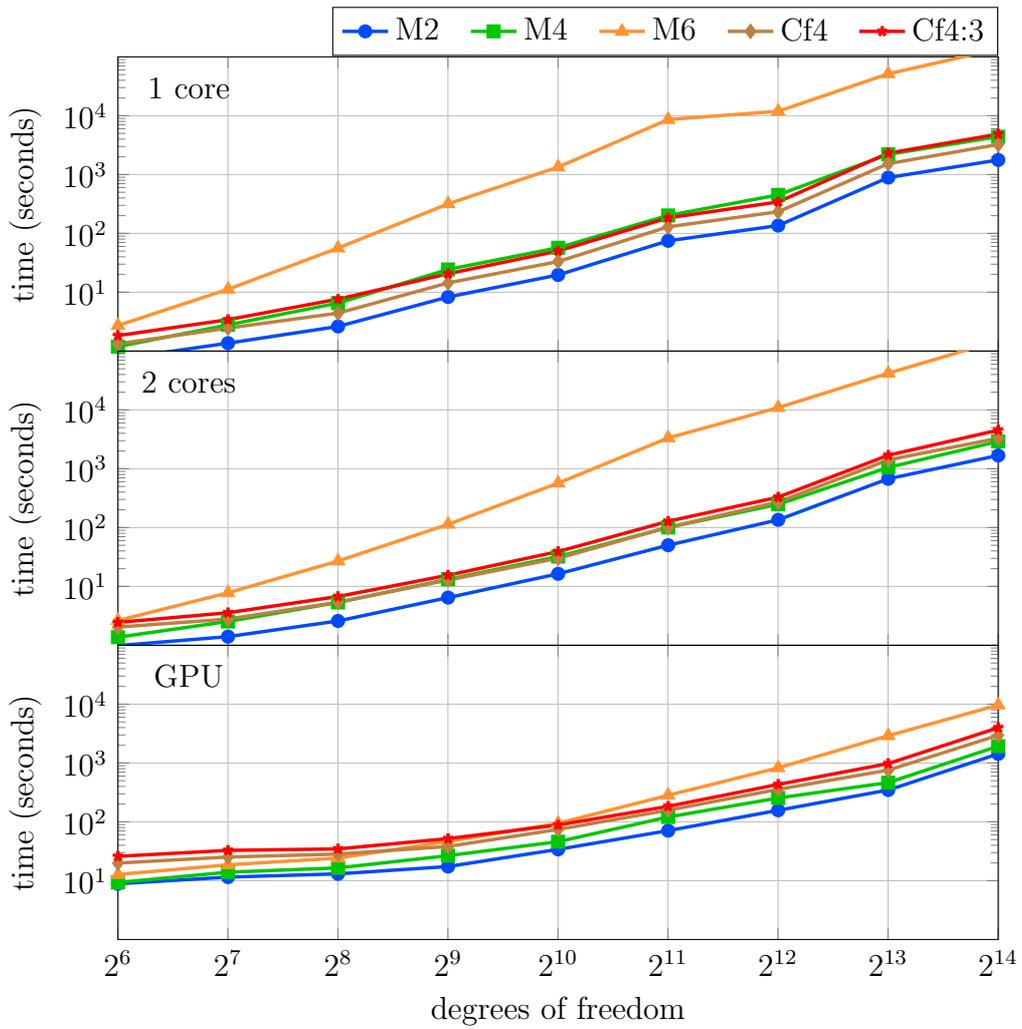
The second experiment, in Figure~\ref{fig:sparse_cvdf_methods}, compares 
    how the methods perform \textcolor{black}{in terms of cost per timestep} for each implementation. 
    The M6 method is clearly outperformed by the \textcolor{black}{four} other methods,
    regardless of the implementation. \textcolor{black}{We remark, however, that for the GPU implementation this gap in performance is significantly smaller.}

\begin{figure}[htb]
\pgfplotsset{height=0.4\linewidth,width=0.96\linewidth,compat=1.10,
    every axis/.append style={legend style={
        /tikz/every even column/.append style={column sep=9pt}}}}
\tikzsetnextfilename{df_vs_t_dense_vs_sparse}
\begin{tikzpicture}[scale=1]%
     \begin{customlegend}[legend style={at={(11.58,4.6)}},
     legend columns=0, 
     legend style={/tikz/every even column/.append style={column sep=9pt}}, 
     legend entries={1 core, multi core, GPU, dense, sparse}]
        \addlegendimage{mblue,line width=1pt, mark=*, only marks}
        \addlegendimage{mgreen,line width=1pt, mark=square*, only marks}
        \addlegendimage{morange,line width=1pt, mark=triangle*, only marks}
        \addlegendimage{black,line width=1pt, dashed, sharp plot}
        \addlegendimage{black,line width=1pt, solid,sharp plot}
    \end{customlegend}
    \begin{loglogaxis}[legend style={at={(1.00,1.17)}}, 
    legend columns=0,cycle list name=long_list, 
    log basis x={2},
    xmin=64, xmax=16384,  grid=major,ymin=1, ymax=1e+5, 
    ytick={10,100,1000,10000},
    xlabel={degrees of freedom}, ylabel={time (seconds)}]
        \node[] at (axis cs:80,3e+4){M4};
        \addplot table[header=false, row sep=\\]{64  1.191079\\128  2.785111\\
          256  6.540586\\512  24.383690\\1024  57.021210\\2048  201.901382\\
          4096 450.884369\\8192 2204.968994\\16384 4395.212891\\};
        \addplot table[header=false, row sep=\\]{64 1.372011\\128 2.535856\\
          256 5.338428\\512 13.220507\\1024 32.191692\\2048 100.522980\\
          4096 248.119659\\8192 1051.603882\\16384 2912.950684\\};
        \addplot table[header=false, row sep=\\]{64  9.311101\\128  13.936012\\
          256  16.465527\\512  26.607653\\1024  45.978233\\2048  120.764725\\
          4096  254.435654\\8192 464.477661\\16384 1931.938965\\};
        \addplot table[header=false, row sep=\\]{64  1.812153\\128  8.166358\\
          256  41.330769\\512  262.967865\\1024  2006.562744\\2048  14311.230469\\
          4096  106015.195312\\};
        \addplot table[header=false, row sep=\\]{64  3.263426\\128  6.636621\\
          256  25.270672\\512  108.441002\\1024  618.947144\\2048 5067.510254\\
          4096 32953.351562\\};
        \addplot table[header=false, row sep=\\]{64  13.836329\\128  22.768105\\
          256  39.954056\\512  81.545128\\1024  163.225052\\2048  426.983032\\
          4096  1626.029419\\};
    \end{loglogaxis}
\end{tikzpicture}
\caption{\label{fig:sparse_densevssparse}
Computational cost vs.\ degrees of freedom for three different 
implementations of the integrator M4. 
The data correspond to Example~\ref{eg:sparse} the solid lines correspond to 
the sparse implementation and the full lines to the dense representation of the
sparse matrix. The  
time step size is fixed to $\tau=10^{-3}$ for computing $\psi(1)$. The
dense computation stops at $2^{12}$ due to storage limitations.}
\end{figure}
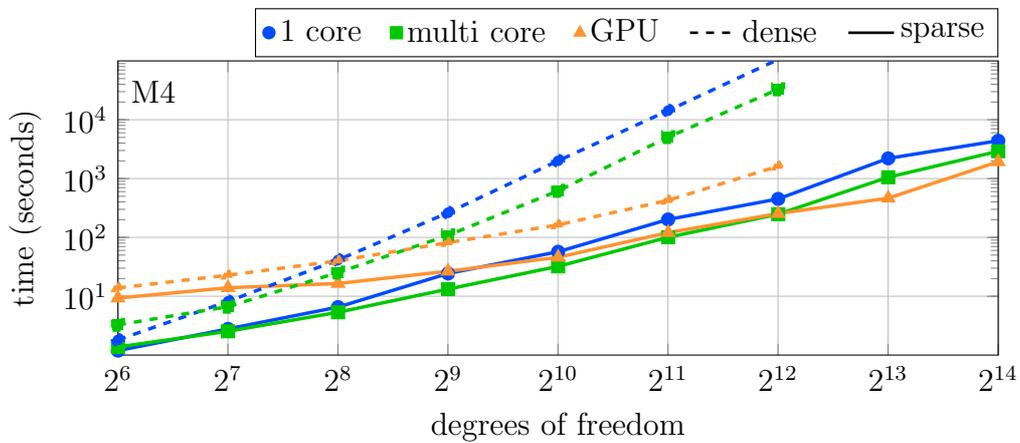

As a consistency check, we show that even though the parallelization of the sparse
matrices is not as beneficial, compared to the dense case, overall sparse algorithms still pay off significantly. This can be clearly seen in Figure~\ref{fig:sparse_densevssparse}. Note, however, that the difference between the two implementations is significantly smaller for the GPU implementation compared to both CPU implementations. Of course, a further advantage of the sparse algorithms are that they consume less memory and we are thus able to solve larger problems.

Finally, we compare the cost of each method, for each of the parallelization techniques, to achieve a specified tolerance. The matrix size is fixed to $N=2^{12}$. The results are shown in Figure \ref{fig:sparse_avsc}. We observe that in this situation the CPU implementation of the \textcolor{black}{Cf4:3} method gives the best result overall. It is also interesting to note that the performance of Cf4 and M4 are very similar for the GPU implementation.

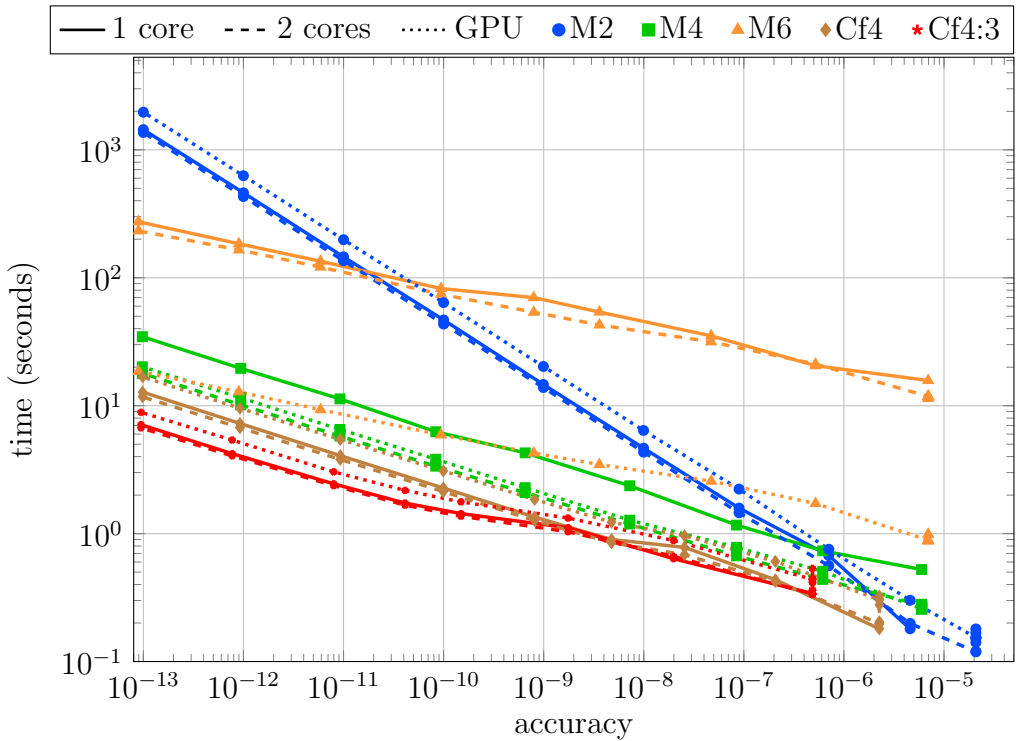
\begin{figure}[htb]
\pgfplotsset{height=0.7\linewidth,width=0.96\linewidth,compat=1.10,
    every axis/.append style={legend style={
        /tikz/every even column/.append style={column sep=9pt}}}}
\tikzsetnextfilename{tol_vs_cpu_sparse}
\begin{tikzpicture}[scale=1]%
    \begin{customlegend}[legend style={at={(11.58,8.7)}},
    legend columns=0, 
    legend style={/tikz/every even column/.append style={column sep=9pt}}, 
    legend entries={1 core, 2 cores, GPU, M2, M4, M6, Cf4, Cf4:3}]
       \addlegendimage{black,line width=1pt, solid, sharp plot}
       \addlegendimage{black,line width=1pt, dashed, sharp plot}
       \addlegendimage{black,line width=1pt, dotted, sharp plot}
       \addlegendimage{mblue,line width=1pt, mark=*, only marks}
       \addlegendimage{mgreen,line width=1pt, mark=square*, only marks}
       \addlegendimage{morange,line width=1pt, mark=triangle*, only marks}
       \addlegendimage{brown,line width=1pt, mark=diamond*, only marks}
       \addlegendimage{mred,line width=1pt, mark=star, only marks}
   \end{customlegend}
    \begin{loglogaxis}[legend style={at={(1.00,1.17)}}, 
    legend columns=0,cycle list name=long_list2, 
    xmin=8e-14, xmax=5e-5,  grid=major,ymin=1e-1, 
    xlabel={accuracy}, ylabel={time (seconds)},
    mark options={solid,scale=0.7}]
      \addplot table[header=false, row sep=\\]{4.5891e-06 1.8025e-01\\ 
        7.0576e-07 6.7436e-01\\ 8.9619e-08 1.5832e+00\\ 9.9521e-09 4.6568e+00\\ 
        9.9239e-10 1.4653e+01\\ 9.9990e-11 4.6803e+01\\ 9.9989e-12 1.4572e+02\\ 
        9.9987e-13 4.6144e+02\\ 9.9986e-14 1.4386e+03\\ };
      \addplot table[header=false, row sep=\\]{5.9789e-06 5.2408e-01\\ 
        6.1405e-07 7.3400e-01\\ 8.4562e-08 1.1662e+00\\ 7.2438e-09 2.3653e+00\\ 
        6.5328e-10 4.2792e+00\\ 8.3560e-11 6.2580e+00\\ 9.2135e-12 1.1306e+01\\ 
        9.3886e-13 1.9572e+01\\ 9.7593e-14 3.4623e+01\\ };  
      \addplot table[header=false, row sep=\\]{6.9472e-06 1.5735e+01\\ 
        5.2062e-07 2.0546e+01\\ 4.7160e-08 3.5142e+01\\ 3.6054e-09 5.3931e+01\\ 
        7.9569e-10 6.9984e+01\\ 9.4271e-11 8.2321e+01\\ 5.8916e-12 1.3458e+02\\
        9.0011e-13 1.8442e+02\\ 8.9451e-14 2.7410e+02\\ };
      \addplot table[header=false, row sep=\\]{2.2584e-06 1.8008e-01\\ 
        2.0663e-07 4.3614e-01\\ 2.5521e-08 7.8099e-01\\ 4.7624e-09 9.0000e-01\\ 
        8.1159e-10 1.3466e+00\\ 9.9354e-11 2.2711e+00\\ 9.2762e-12 4.0717e+00\\ 
        9.2318e-13 7.2638e+00\\ 9.8255e-14 1.2776e+01\\ };
      \addplot table[header=false, row sep=\\]{4.8521e-07 3.3440e-01\\ 
        4.8521e-07 3.3497e-01\\ 4.8522e-07 3.6058e-01\\ 4.8522e-07 3.3717e-01\\ 
        4.8522e-07 3.3840e-01\\ 4.8522e-07 3.4004e-01\\ 1.9836e-08 6.3705e-01\\ 
        1.7530e-09 1.1117e+00\\ 1.4904e-10 1.4343e+00\\ 4.1216e-11 1.7331e+00\\ 
        7.9779e-12 2.4498e+00\\ 7.7185e-13 4.2179e+00\\ 9.4785e-14 7.1021e+00\\ };
      \addplot table[header=false, row sep=\\]{2.08218e-05 1.6757e-01\\ 
        2.08218e-05 1.4125e-01\\ 2.08218e-05 1.2001e-01\\ 2.08218e-05 1.1904e-01\\ 
        4.58914e-06 1.9878e-01\\ 7.05762e-07 5.6060e-01\\ 8.96197e-08 1.4560e+00\\ 
        9.95217e-09 4.3355e+00\\ 9.92397e-10 1.3857e+01\\ 9.99903e-11 4.3477e+01\\ 
        9.99893e-12 1.3576e+02\\ 9.99876e-13 4.3107e+02\\ 9.99865e-14 1.3650e+03\\ };
      \addplot table[header=false, row sep=\\]{5.979e-06 2.749e-01\\ 
        5.978e-06 2.682e-01\\ 5.978e-06 2.697e-01\\ 5.978e-06 2.745e-01\\ 
        5.978e-06 2.753e-01\\ 6.140e-07 4.386e-01\\ 8.456e-08 6.704e-01\\ 
        7.243e-09 1.188e+00\\ 6.532e-10 2.091e+00\\ 8.356e-11 3.365e+00\\ 
        9.213e-12 5.781e+00\\ 9.388e-13 1.016e+01\\ 9.759e-14 1.788e+01\\ };  
      \addplot table[header=false, row sep=\\]{6.9472e-06 1.1391e+01\\ 
        6.9472e-06 1.1861e+01\\ 6.9472e-06 1.1850e+01\\ 6.9472e-06 1.1738e+01\\ 
        6.9472e-06 1.1813e+01\\ 5.2062e-07 2.1143e+01\\ 4.7160e-08 3.1566e+01\\ 
        3.6054e-09 4.2781e+01\\ 7.9569e-10 5.3754e+01\\ 9.4271e-11 7.4307e+01\\ 
        5.8916e-12 1.2117e+02\\ 9.0011e-13 1.6517e+02\\ 8.9451e-14 2.3269e+02\\};
      \addplot table[header=false, row sep=\\]{2.2584e-06 2.7715e-01\\ 
        2.2584e-06 1.9711e-01\\ 2.2584e-06 2.0442e-01\\ 2.2584e-06 1.9900e-01\\ 
        2.2584e-06 1.9869e-01\\ 2.0663e-07 4.2292e-01\\ 2.5521e-08 6.8403e-01\\ 
        4.7624e-09 8.4841e-01\\ 8.1159e-10 1.2656e+00\\ 9.9354e-11 2.1303e+00\\ 
        9.2762e-12 3.8023e+00\\ 9.2318e-13 6.7232e+00\\ 9.8255e-14 1.1718e+01\\ };
      \addplot table[header=false, row sep=\\]{4.8521e-07 5.3446e-01\\ 
        4.8521e-07 4.1372e-01\\ 4.8522e-07 4.1702e-01\\ 4.8522e-07 3.6187e-01\\ 4.8522e-07 3.3597e-01\\ 4.8522e-07 3.3719e-01\\ 1.9836e-08 6.6461e-01\\ 1.7530e-09 1.0343e+00\\ 1.4904e-10 1.3674e+00\\ 4.1216e-11 1.6686e+00\\ 7.9779e-12 2.3725e+00\\ 7.7185e-13 4.0521e+00\\ 9.4785e-14 6.7058e+00\\};
      \addplot table[header=false, row sep=\\]{2.0821e-05 1.7999e-01\\ 
        2.0821e-05 1.5326e-01\\ 2.0821e-05 1.5367e-01\\ 2.0821e-05 1.5411e-01\\ 
        4.5891e-06 3.0141e-01\\ 7.0576e-07 7.5509e-01\\ 8.9619e-08 2.2304e+00\\ 
        9.9521e-09 6.4032e+00\\ 9.9239e-10 2.0266e+01\\ 9.9990e-11 6.3927e+01\\ 
        9.9989e-12 1.9840e+02\\ 9.9989e-13 6.2710e+02\\ 9.9986e-14 1.9683e+03\\};
      \addplot table[header=false, row sep=\\]{5.9789e-06 2.8076e-01\\ 
        5.9789e-06 2.5802e-01\\ 5.9789e-06 2.5727e-01\\ 5.9789e-06 2.5778e-01\\ 
        5.9789e-06 2.5797e-01\\ 6.1405e-07 5.0700e-01\\ 8.4562e-08 7.8159e-01\\ 
        7.2438e-09 1.2742e+00\\ 6.5328e-10 2.2885e+00\\ 8.3560e-11 3.8149e+00\\ 
        9.2135e-12 6.5205e+00\\ 9.3886e-13 1.1534e+01\\ 9.7593e-14 2.0154e+01\\ };  
      \addplot table[header=false, row sep=\\]{6.9472e-06 9.8838e-01\\ 
        6.9472e-06 8.7776e-01\\ 6.9472e-06 8.7989e-01\\ 6.9472e-06 8.8315e-01\\ 
        6.9472e-06 8.8963e-01\\ 5.2062e-07 1.7202e+00\\ 4.7160e-08 2.5742e+00\\ 
        3.6054e-09 3.4667e+00\\ 7.9569e-10 4.2644e+00\\ 9.4271e-11 5.9118e+00\\ 
        5.8916e-12 9.3055e+00\\ 9.0011e-13 1.2835e+01\\ 8.9451e-14 1.8549e+01\\ };
      \addplot table[header=false, row sep=\\]{2.2584e-06 3.2460e-01\\ 
        2.2584e-06 3.0377e-01\\ 2.2584e-06 3.0467e-01\\ 2.2584e-06 3.0427e-01\\ 
        2.2584e-06 3.0466e-01\\ 2.0663e-07 6.0571e-01\\ 2.5521e-08 9.6300e-01\\ 
        4.7624e-09 1.2283e+00\\ 8.1159e-10 1.8472e+00\\ 9.9354e-11 3.0890e+00\\ 
        9.2762e-12 5.4429e+00\\ 9.2318e-13 9.5421e+00\\ 9.8255e-14 1.6793e+01\\ };
      \addplot table[header=false, row sep=\\]{4.8529e-07 4.4432e-01\\ 
        4.8521e-07 4.3683e-01\\ 4.8522e-07 4.3682e-01\\ 4.8522e-07 4.3710e-01\\ 
        4.8522e-07 4.3852e-01\\ 4.8522e-07 4.3897e-01\\ 1.9836e-08 8.8348e-01\\ 
        1.7530e-09 1.3240e+00\\ 1.4904e-10 1.7706e+00\\ 4.1216e-11 2.1760e+00\\ 
        7.9779e-12 3.0394e+00\\ 7.7185e-13 5.3884e+00\\ 9.4785e-14 8.8723e+00\\ };
    \end{loglogaxis}
\end{tikzpicture}
\caption{\label{fig:sparse_avsc}%
Accuracy vs.\ computational cost for the five integrators M2, M4, 
M6, Cf4, and Cf4:3 when using different parallelization techniques. 
The date corresponds to Example~\ref{eg:sparse} with fixed 
matrix dimension $N=2^{12}$. For a fixed step size $\tau$ the error of $\psi(1)$
is measured.}
\end{figure}

\end{bsp}

\section{Conclusion \label{sec:conclusion}}

In this work we have investigated whether parallelization on GPUs is a viable option for Magnus integrators based on Leja interpolation. The numerical solutions have been computed in the context of the Schr\"odinger equation using a local and a non-local Heisenberg model. Implementing these algorithms on modern multi-core CPU and GPU systems shows that in the dense case the GPU implementation is able to achieve speed-ups of up to a factor of $10$. In the sparse case only very large problems benefit from GPU acceleration and the achieved speed-ups are more modest. 

From a numerical analysis point of view, we observe that the commutator-free fourth order method (Cf4 \textcolor{black}{and, in particular, Cf4:3}) is usually the best choice. Although for both multi-core CPUs and GPUs the traditional fourth order exponential integrator (M4) is usually quite competitive. This is in stark contrast to the sequential case, where M4 is significantly slower compared to the Cf4 method. This behavior is due to the fact that the matrix-matrix products required to compute the commutators can be more efficiently parallelized. In fact, the per time step cost of Cf4 \textcolor{black}{and Cf4:3} is even larger compared to M4. \textcolor{black}{That is, the increased efficiency is due to the increased accuracy (and certainly not due to a decrease in cost, which is often given as a motivation for constructing such methods)} This has important implications for future research in constructing more efficient Magnus type integrators, as modern hardware calls into question the design philosophy of eliminating matrix-matrix products at all costs (at least this is true for the problem sizes currently accessible on a workstation).

\section*{Acknowledgement}
This work was supported by the Tiroler Wissenschaftsfonds (TWF) under grant number UNI-0404/1531.

\bibliography{magnus-gpu}
\bibliographystyle{plainnat}
\end{document}